%% file: inoue.tex
\documentclass[dvipdfmx]{PoS}
\usepackage{url}

\title{Nuclear physics from QCD on lattice}

\ShortTitle{Nuclear phyics form QCD on lattice}

\author{\speaker{Takashi Inoue}\thanks{A footnote may follow.}\\
        Nihon University, College of Bioresource Sciences, Kanagawa 252-0880, Japan\\
        E-mail: \email{inoue.takashi@nihon-u.ac.jp}}

\author{for HAL QCD Collaboration}

\abstract{
\input{abst}
}

\FullConference{The 8th International Workshop on Chiral Dynamics, CD2015 ***\\
		29 June 2015 - 03 July 2015\\
		Pisa,Italy}

\begin{document}

\def \etal{{\it et al.\,}}
\def \etc{{\it etc.\,}}
\def \ie{{\it i.e.\,}}
\def \eg{{\it e.g.\,}}

\input{section_1}
\input{section_2}
\input{section_3}
\input{section_4}
\input{section_5}
\input{section_6}
\input{section_7}

\section*{Acknowledgements}
The author thank the maintainer of CPS++~\cite{CPS} whose modified version is used in this work,
and PACS-CS collaboration for providing their DDHMC/PHMC code,
and JLDG/ILDG team~\cite{JLDG_ILDG} for providing storage to save our lattice QCD data. 
Numerical computations of this work were carried out at Univ.~of Tsukuba supercomputer system (T2K).
This research is supported in part by the JSPS Grant-in-Aid for Scientific Research (C)26400281.

\end{document}

%% file: abst.tex
We have presented a strategy to study nuclei and nuclear matters from first principles, namely, from QCD.
We first compute nucleon-nucleon potentials numerically in lattice QCD,
and then use them to investigate properties of nuclei and nuclear matter by various methods developed in nuclear physics.
As a demonstration that this strategy works, mass and structure of $^{4}$He, $^{16}$O and $^{40}$Ca,
and equation of state of nuclear matters are determined with the lattice QCD induced two-nucleon potentials in a heavy quark region as an input.
We have found that these nuclei and the symmetric nuclear matter are bound
at one quark mass corresponding to the pseudo-scalar meson (pion) mass of 469 MeV (the octet baryon (nucleon) mass of 1161 MeV).
The obtained binding energy per nucleon has a uniform mass-number $A$ dependence
which is consistent to the Bethe-Weizs{\"a}cker mass formula qualitatively.
The present study demonstrates that our strategy works well to investigate various properties of atomic nuclei
and nuclear matter starting from QCD, without depending on models or experimental information about the nuclear force.

%% file: section_1.tex
\section{Introduction}
\label{sec:intro}

Nuclear theory has been developed extensively since the 1930's~\cite{RingSchuck}.
It started from the liquid drop model and the empirical mass formula.
Then, the nuclear shell model, supported by mean field theory and Brueckner's theory, achieved lots of success.
Today, variational methods with some advanced technique can provide us exact solution
for light nuclei~\cite{Hiyama:2003cu,Varga:1995dm,Viviani:2004vf}.
Several sophisticated theories and models are developed for heavier nuclei
over these decades~\cite{Hagen:2007hi,Fujii:2009bf,Dickhoff:2004xx,Pieper:2001mp,Navratil:2000ww,Shimizu:2012mv}.
However, these theoretical studies need to use input data from experiment.

It has been a long time since QCD was established as the fundamental theory of the strong interaction.
In principle, we are able to explain everything from QCD, including hadron spectrum, hadron structure,
binding energy of nuclei, and so on.
However, it is difficult to do because of the non-perturbative nature of QCD.
Thanks to the recent advances in lattice QCD, masses of the ground state hadrons are reproduced well~\cite{Aoki:2009ix,Durr:2010aw},
and structure of hadrons are going to be reproduced~\cite{Alexandrou:2011db}.
However, explaining properties of nuclei and nuclear matters starting from QCD still remains one of the most challenging problem in physics.

To study nuclei based on QCD, today's most popular approaches are ones based on the chiral Lagrangian.
The Lagrangian is constructed using the chiral symmetry of QCD with vanishing quark mass,
and its coefficients (low energy constants) are determined by fitting to experimental data.
In most case, potentials of two-nucleon ($NN$) and three-nucleon ($NNN$) forces are obtained
in a perturbation theory with the particular power counting~\cite{Weinberg:1990rz},
and then potentials are applied to the nuclear theories~\cite{Gezerlis:2013ipa}.
Sometimes, the Lagrangian is studied directly by solving the Bethe-Salpeter equation for small nuclei~\cite{Konig:2011yq}.
Recently, the Lagrangian has been studied numerically on the lattice~\cite{Epelbaum:2009pd}.
These theoretical studies are partly based on QCD but need experimental input.

There are several pioneering attempts to extract scattering observables of two-nucleon system and binding energy of light nuclei
from lattice QCD numerical simulations~\cite{Yamazaki:2009ua,Beane:2011iw}.
However, it seems that there is a fundamental difficulty in direct extraction of energy shift of multi-baryon system in lattice QCD.
Namely, it is difficult or practically impossible to achieve the ground state saturation of the corresponding correlation function (plateau crisis).
The origin of this difficulty is the following.
In order to study multi-baryon system in lattice QCD, one needs to take spacial volume of lattice sufficiently large. 
When spacial volume is large, intervals between energy levels of discretized continuum become small.
Moreover, excitation energy of nucleus is much smaller than that of hadron in the first place. 
Therefore, it is very hard to suppress excited state contribution to the correlation function. 
This difficulty is so fundamental that we take completely different approach which does not require the ground state saturation.
Our approach consists of two stages.
First, we extract potential of interaction between hadrons in lattice QCD numerical simulation.
Then, we solve the Schr{\"o}dinger equation involving the potential and obtain physical observables of multi-hadron system interested.
We have found that this approach is feasible and promising.

This paper is organized as follows.
In section 2, we explain the method to extract hadron-hadron interaction in lattice QCD simulation.
In section 3, we present our simulation setup and obtain two-nucleon potentials.
In section 4, we apply the potentials to the light nucleus $^4$He.
In section 5, we apply the potentials to the medium-heavy nuclei $^{16}$O and $^{40}$Ca.
In section 6, we apply the potentials to the infinite nuclear matters.
Section 7 is devoted to summary and discussion.

%% file: section_2.tex
\section{Nuclear force from QCD}
\label{sec:hal}

In 2006, Ishii \etal proposed a method to extract nucleon-nucleon ($N\!N$) interaction
from QCD on lattice~\cite{Ishii:2006ec}.
This method has been applied to many others systems successfully~\cite{Nemura:2008sp},
and called as HAL QCD method, today.
This method utilizes the equal-time Nambu-Bethe-Salpeter (NBS) wave function
which is defined for the two-nucleon case by
\begin{equation}
 \varphi_{\vec{k}}(\vec r)  = \sum_{\vec x} 
   \langle 0 \vert N(\vec x + \vec r,0)N(\vec x,0)\vert  {N} {N}, \vec{k} \rangle
\label{eqn:psi}
\end{equation}
where $\vert N N, \vec{k}\rangle$ is a two-nucleon QCD eigenstate in the rest frame
with a relative momentum $\vec{k}$ and $N(\vec x,t)$ is the nucleon field operator.
With the NBS wave function, a non-local potential $U(\vec r, {\vec r\,}')$,
can be defined though a Schr\"{o}dinger type equation as
\begin{equation}
  -\frac{\nabla^2}{2\mu} \, \varphi_{\vec{k}}(\vec r) 
 ~+~ \int \!\! d^3 {\vec r\,}' \, U(\vec r, {\vec r\,}') \, \varphi_{\vec{k}}({\vec r\,}')
 ~=~ E_{\vec k} \, \varphi_{\vec{k}}(\vec r)
\label{eqn:defofu}
\end{equation}
where $E_{\vec k}\,=\,\frac{\vec{k}^2}{2\mu}$ with the reduced mass $\mu\,=\,\frac{M_{N}}{2}$.
Note that the potential $U(\vec r, {\vec r\,}')$ is defined as common
for all energy eigenstates (for all $\vec{k}$) below inelastic threshold.

On the other hand, in lattice QCD numerical simulations, one can measure the 4-point correlation function
defined for the two-nucleon case by
\begin{equation}
\Psi(\vec{r}, t) \equiv  \sum_{\vec{x}}\langle 0 \vert N(\vec x + \vec r,t)N(\vec x,t) {\cal J}(t_0)\vert 0 \rangle
\end{equation}
where ${\cal J}(t_0)$ is a source operator which creates two-nucleon states at $t_0$.
By inserting the complete set between $N(\vec x,t)$ and ${\cal J}(t_0)$, 
one can see that this correlation function contains the NBS wave function $\varphi_{\vec{k}}(\vec r)$ as
\begin{equation}
\Psi(\vec{r}, t) = \sum_{\vec{k}} \, A_{\vec{k}} \, \varphi_{\vec{k}}(\vec{r}) \, e^{-W_{\vec{k}}(t-t_0)} ~+~ \cdots 
\end{equation}
with the normalization $A_{\vec k} \,=\, \langle NN,\vec{k} \vert {\cal J}(t_0) \vert 0 \rangle\,$,
the total energy $W_{\vec{k}} \,=\, 2 \sqrt{M_N^2 + \vec{k}^2} \,\simeq\, 2 M_N + E_{\vec k}\,$,
and ellipsis denotes inelastic contributions, which can be ignored for reasonably large $t-t_0$.

Because the equation (\ref{eqn:defofu}) is linear in the NBS wave function $\varphi_{\vec{k}}(\vec r)$, 
and the potential $U(\vec r, {\vec r\,}')$ is common for all $\vec{k}$, one easily obtains an equation
\begin{equation}
     \left[ 2 M_N ~-~ \frac{\nabla^2}{2\mu} \right] \, \Psi(\vec r, t)
   ~+~ \int \!\! d^3 {\vec r\,}' \, U(\vec r, {\vec r\,}') \, \Psi({\vec r\,}', t)
   ~=~ - \frac{\partial}{\partial t} \, \Psi(\vec r, t)
\label{eqn:eqforu}
\end{equation}
which relates $\Psi(\vec{r}, t)$ and $U(\vec r, {\vec r\,}')$.
On can use this equation to extract interaction potentials from lattice QCD data.
It was shown that this equation makes the extraction very stable and robust. We show some examples below.

Because available lattice QCD data are limited usually, extracting a non-local potential is not practical.
Therefor, in our actual studies, we apply the velocity (derivative) expansion of the non-local potential
\begin{equation}
U(\vec r,\vec r') ~=~ \delta^{3}(\vec r-\vec r')V(\vec r,\nabla)
~=~ \delta^{3}(\vec r-\vec r')\left\{ V_0(\vec r) ~+~ O(\nabla) \right\}
\end{equation}
and truncate higher order derivative terms. 
When we truncate, the leading order potential $V_0(\vec r)$ is obtained,
from the equation (\ref{eqn:eqforu}), by
\begin{equation}
  V_0(\vec r) = \frac{1}{2\mu}\frac{\nabla^2 \Psi(\vec r, t)}{\Psi(\vec r, t)} ~-~ 
                \frac{\frac{\partial}{\partial t} \Psi(\vec r, t)}{\Psi(\vec r, t)} ~-~ 2 M_N ~.
\label{eq:vr}
\end{equation}
This can be rewritten in a more convenient and statistically advantageous form
\begin{equation}
  V_0(\vec r) = \frac{1}{2\mu}\frac{\nabla^2 R(\vec r, t)}{R(\vec r, t)} ~-~ 
                \frac{\frac{\partial}{\partial t} R(\vec r, t)}{R(\vec r, t)}
\label{eq:vr2}
\end{equation}
where $R(\vec r, t)$ is defined by $R(\vec r, t) = {\Psi(\vec r, t)}/{B(t)^2}$
with the single hadron 2-point function $B(t)$.

\begin{figure}[t]
\centering
\includegraphics[width=0.45\textwidth]{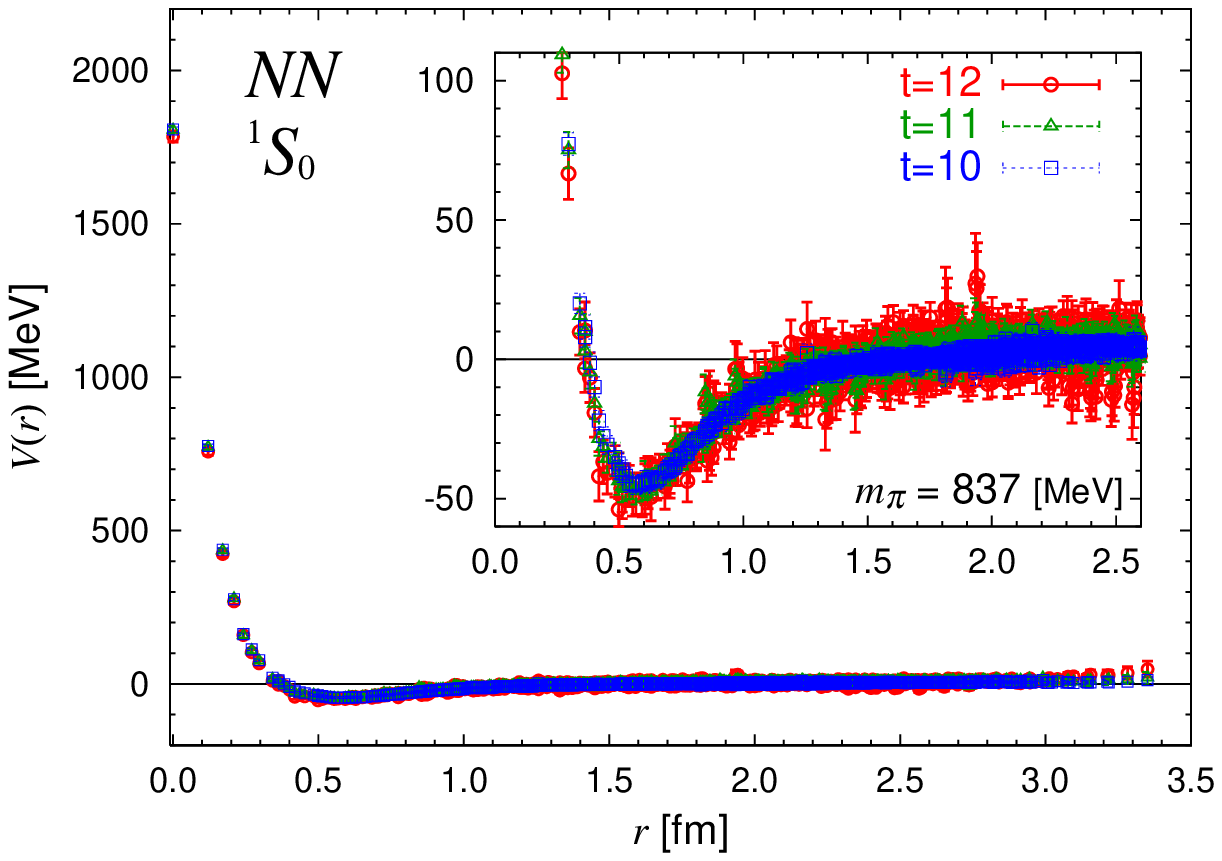}~~
\includegraphics[width=0.45\textwidth]{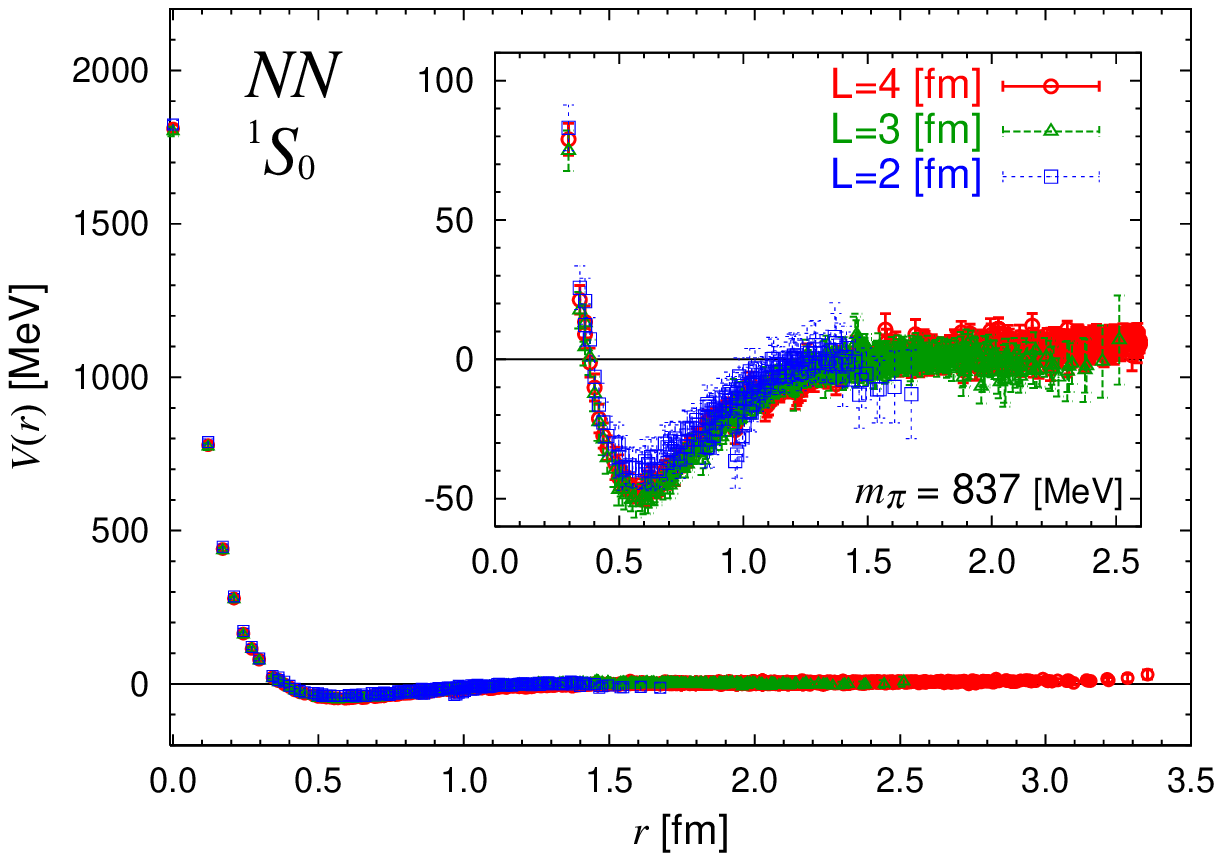}
\caption{Potential of $NN$ interaction in $^1S_0$ partial wave extracted from lattice QCD at pion mass 837 MeV.
Left panel shows ones extracted from data at three different time slices $t=10$, 11, and 12.
Right panel show ones measured on three different size of lattice $L=2$, 3 and 4 fm.}
\label{fig:robust}
\end{figure} 

It is important to note that the equations (\ref{eqn:eqforu}) and (\ref{eq:vr}) (or (\ref{eq:vr2}) ) do NOT require
the ground state saturation for $\Psi(\vec r, t)$, which is usually very difficult or almost impossible to achieve
in actual lattice QCD numerical simulations, in particular on a large spacial volume for two-baryon systems.
In fact, extracted potentials are independent of $\,t-t_0\,$ in this method,
as long as $\,t-t_0\,$ is large enough so that a single hadron $B(t)$ is saturated by its ground state.
Fig.~\ref{fig:robust} shows potential of $NN$ interaction in $^1S_0$ partial wave extracted from lattice QCD at pion mass 837 MeV. 
There, the wall type quark source ${\cal J}(t_0)$ is placed at origin of time axis, namely $\,t_0=0\,$.
In the left panel, ones extracted from data of $\Psi(\vec r, t)$ at three different time slices $t=10$, 11, and 12 are shown.
The used data of $\Psi(\vec r, t)$ are measured on relatively large volume ($L=4$ fm),
and hence not saturated by the ground state at all at around $t=10$ to 12, and depend on $\,t\,$ essentially.
Nevertheless, extracted potentials are independent on $\,t\,$ as we can see explicitly in the figure.
This is an example which shows that the HAL QCD method provides a crucial solution
to the plateau crisis in study of muluti-hadron systems in lattice QCD. 

It is also remarkable that the potential is independent on the spacial volume of lattice,
as long as size of lattice is larger than the largest range of interaction between hadrons.
Fig.~\ref{fig:robust}, in the right panel, shows $NN$ potential in $^1S_0$ partial wave 
extracted from data measured on three different volume with $L=2$, 3, and 4 fm.
We can see that extracted potentials agree each other, except that $L=2$ fm seems a little small.
This agreement means that lattice QCD calculation with one volume is enough in the potential method,
and we do NOT need to do infinite-volume extrapolations which consume a lot of time and money. 
This is a significant advantage of the HAL QCD method over the conventional one.

Once potentials are obtained, physical observables are obtained by solving the Schr\"{o}dinger equation in infinite volume.
More physical observables can be obtained in this approach than the conventional method using energy shift.
For example, we can predict scattering phase shift as a function of energy.
Moreover, we can study the properties of nuclei and infinite nuclear matter.
This is another remarkable advantage of the HAL QCD method.
Note that a direct lattice QCD simulation of heavy nuclei must be formidably expensive
even with the new algorithm for the Wick contraction~\cite{Doi:2012xd,Detmold:2012eu,Gunther:2013xj}.

%% file: section_3.tex
\section{Setup of lattice QCD simulations and resulting two-nucleon potentials}
\label{sec:setup}

\begin{table}[t]
\caption{Lattice parameters such as
the lattice size, the inverse coupling constant $\beta$, the clover coefficient $c_{\rm sw}$,
the lattice spacing $a$ and the physical extension $L$. See ref.~\cite{CPPACS-JLQCD} for details.}
\label{tbl:lattice}
\medskip
\centering
 \begin{tabular}{c|c|c|c|c}
   \hline   \hline
    size             & ~~ $\beta$ ~~ & ~~ $c_{\rm sw}$ ~~ & ~ $a$ [fm] ~ & ~$L$ [fm]~  \\
   \hline 
    $32^3 \times 32$  &   1.83  &   1.761   &  0.121(2)  & 3.87  \\
   \hline  \hline
 \end{tabular}
\end{table}

In general, for lattice QCD numerical simulations with dynamical quarks,
we need gauge configuration ensembles generated beforehand. 
The gauge configuration ensembles at the physical point 
generated by the PACS-CS collaboration~\cite{Aoki:2009ix} and the BMW collaboration~\cite{Durr:2010aw}, 
were intended to study single hadron properties, 
and their spacial volume ($L\simeq 2$ fm) are small even for two-nucleon system.
Therefore, in this study, we use gauge configuration ensembles which we generated
on relatively large spacial volume ($L\simeq 4$ fm) but with un-physical quark masses.
Employed actions are the renormalization group improved Iwasaki gauge action~\cite{Iwasaki:2011np}, 
and the non-perturbatively $O(a)$ improved Wilson quark action.
Our simulation parameters are summarized in Table~\ref{tbl:lattice}.

\begin{table}[t]
\caption{Quark hopping parameter $\kappa_{\rm uds}$ and corresponding hadron masses,
$M_{\rm PS}$, $M_{\rm V}$, $M_{\rm B}$ for pseudo-scalar meson, vector meson and octet baryon, respectively.}
\label{tbl:mass}
\medskip
\centering
 \begin{tabular}{c|c|c|c|c}
   \hline  \hline
    ~~ $\kappa_{\rm uds}$ ~~ & $M_{\rm PS}$ [MeV] &  $M_{\rm V}$ [MeV] &  $M_{\rm B}$ [MeV] &
   $N_{\rm cfg}\,/\,N_{\rm traj}$ \\
   \hline 
     0.13660 &   1170.9(7) &   1510.4(0.9) & 2274(2) & 420\,/\,4200 \\
     0.13710 &   1015.2(6) &   1360.6(1.1) & 2031(2) & 360\,/\,3600 \\
     0.13760 & ~\,836.5(5) &   1188.9(0.9) & 1749(1) & 480\,/\,4800 \\
     0.13800 & ~\,672.3(6) &   1027.6(1.0) & 1484(2) & 360\,/\,3600 \\
     0.13840 & ~\,468.6(7) & ~\,829.2(1.5) & 1161(2) & 720\,/\,3600 \\
   \hline  \hline
 \end{tabular}
\end{table}

In lattice QCD, mass of quarks are tuned by the so called quark hopping parameters $\kappa_{\rm i}$.
We choose $\kappa_{\rm u} = \kappa_{\rm d} = \kappa_{\rm s} = \kappa_{\rm uds}$ in our gauge configuration ensembles.
In other words, we set strange quark mass equal to up and down quark mass.
This is in order to study the flavor $SU(3)$ symmetric world.
The flavor symmetric world is known to be very useful to capture essential features of hadron interaction.
For example, S-wave interaction between two octet-baryons are reduced to six independent interactions.
This advantage is used in lattice QCD studies~\cite{Inoue:2010hs,Beane:2012vq}.
We generated five ensembles with different value of $\kappa_{\rm uds}$. 
The values of $\kappa_{\rm uds}$ and measured hadron masses are given in Table~\ref{tbl:mass}.
One sees that we can study nucleonic systems in lattice QCD at wide range of nucleon mass and pion mass with these ensembles.

We measure the nucleon 2-point functions $B(t)$ and nucleon 4-point functions $\Psi(\vec{r}, t)$.
In our measurement, we use the wall type quark source and the point type nucleon field operator at sink
\begin{eqnarray}
p_{\alpha}(x) &=& +\epsilon_{c_1,c_2,c_3}\,(C\gamma_5)_{\beta_1,\beta_2}\,\delta_{\beta_3,\alpha}\,u(\xi_1)d(\xi_2)u(\xi_3) ,
\\
n_{\alpha}(x) &=& +\epsilon_{c_1,c_2,c_3}\,(C\gamma_5)_{\beta_1,\beta_2}\,\delta_{\beta_3,\alpha}\,u(\xi_1)d(\xi_2)d(\xi_3)
\end{eqnarray}
with $\xi_i = \left(c_i, \beta_i, x \right)$.
We use Dirichlet boundary conditions in the temporal direction.
In order to reduce noise and enhance signal, we measure the functions 16 times for each configuration
by shifting the source in the temporal direction, and then average over sources.
We utilize also an average over forward and backward propagations in time.

\begin{figure}[t]
\begin{center}
\includegraphics[width=0.45\textwidth]{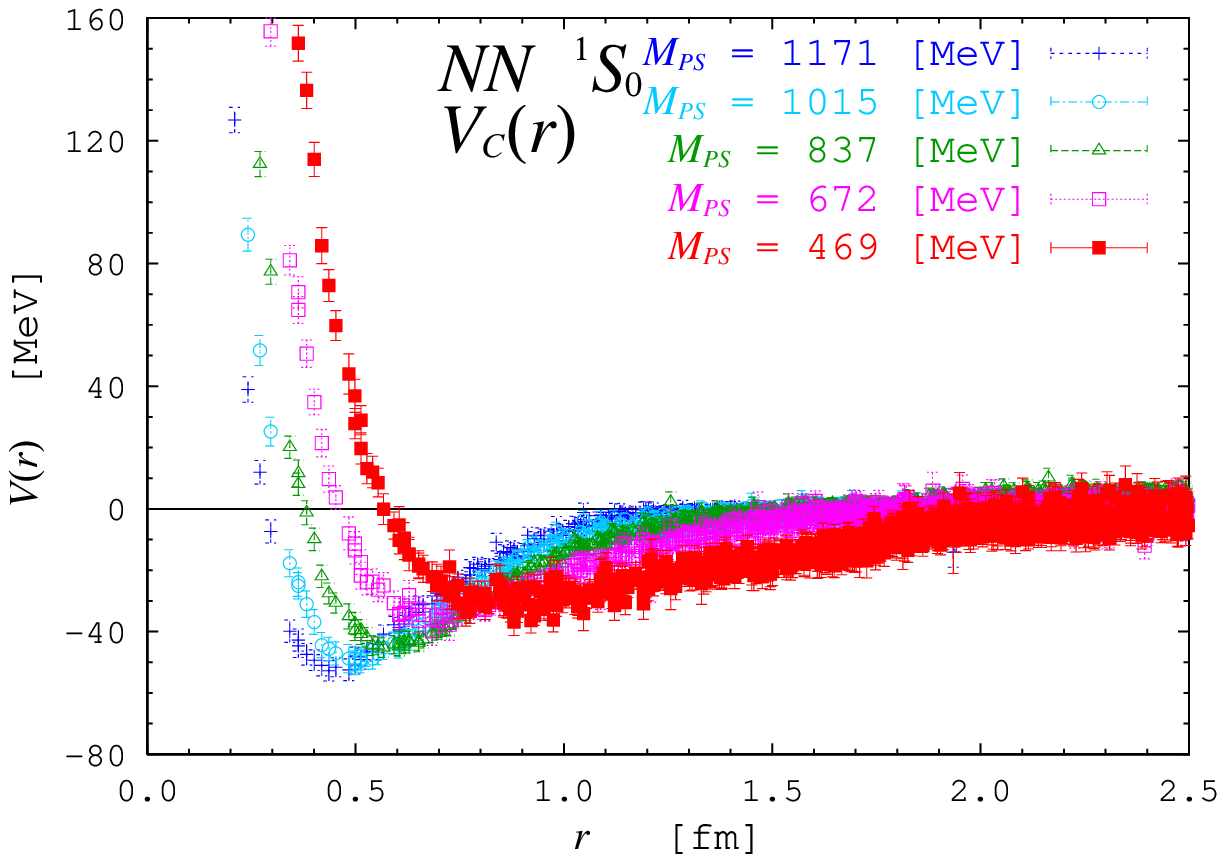} \hspace{63mm} \newline
\includegraphics[width=0.45\textwidth]{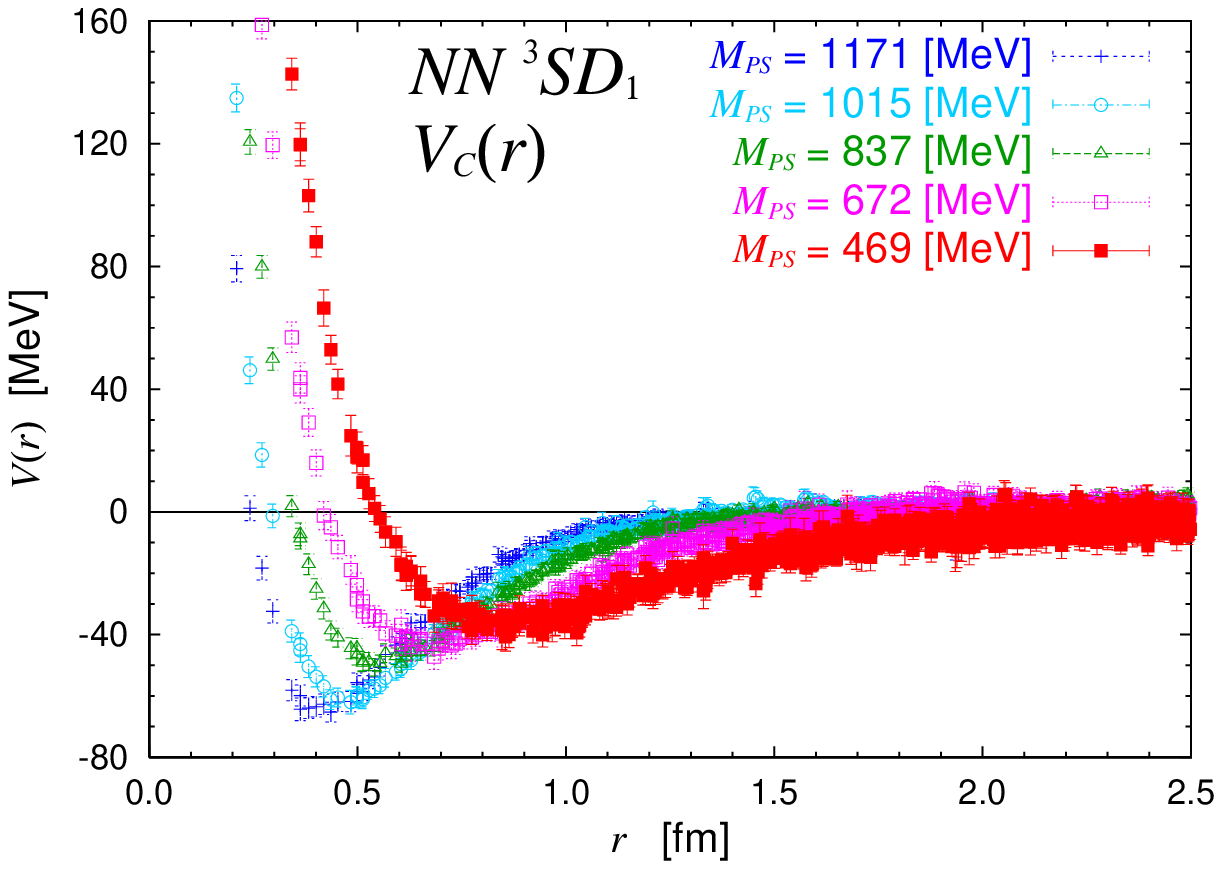}~~
\includegraphics[width=0.45\textwidth]{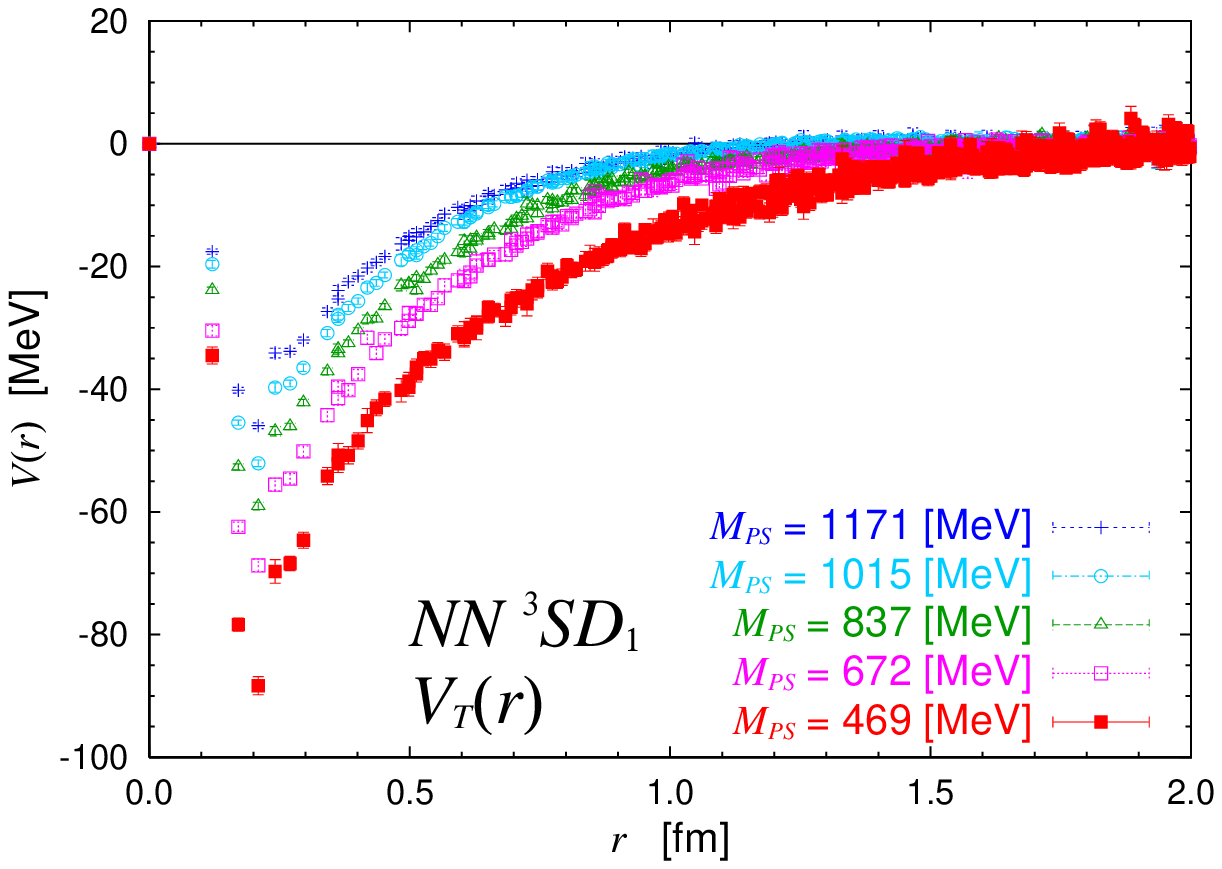}
\end{center}
\caption{Potentials of $NN$ interaction extracted from lattice QCD at five flavor SU(3) symmetric points.}
\label{fig:potNN}
\end{figure} 

Fig.~\ref{fig:potNN} shows potentials of $NN$ interaction extracted from lattice QCD at the five flavor SU(3) symmetric points.
The vertical bars show statistical error estimated in the Jackknife method.
First of all, one sees that the lattice QCD induced $NN$ potentials share common features
with the phenomenological ones (\eg the Argonne V18 potential given in ~ref.~\cite{Wiringa:1994wb}),
namely, a repulsive core at short distance, an attractive pocket at medium distance, and a strong tensor force.
Accordingly, these lattice QCD $NN$ potentials well reproduce the aspects of $N\!N$ scattering observables~\cite{Inoue:2010hs}. 
However, the strength of the lattice QCD nuclear force at the five investigated points, 
is weaker than the empirical one. In particular, the deuteron, \ie the bound state in $^3S_1$-$^3D_1$ channel, is not supported.
This failure is due to the heavy up and down quark in our simulations.
In fact, in Fig.~\ref{fig:potNN}, one see that the lattice QCD nuclear force become stronger and stronger as the degenerate quark mass decreases.
Therefore, one can expect that, when $N\!N$ potentials are extracted from lattice QCD simulation at the physical point,
any two-nucleon observables will be reproduced quantitatively.

\begin{figure}[t]
\begin{minipage}[t]{0.475\columnwidth}
\centering
 \includegraphics[width=1.0\textwidth]{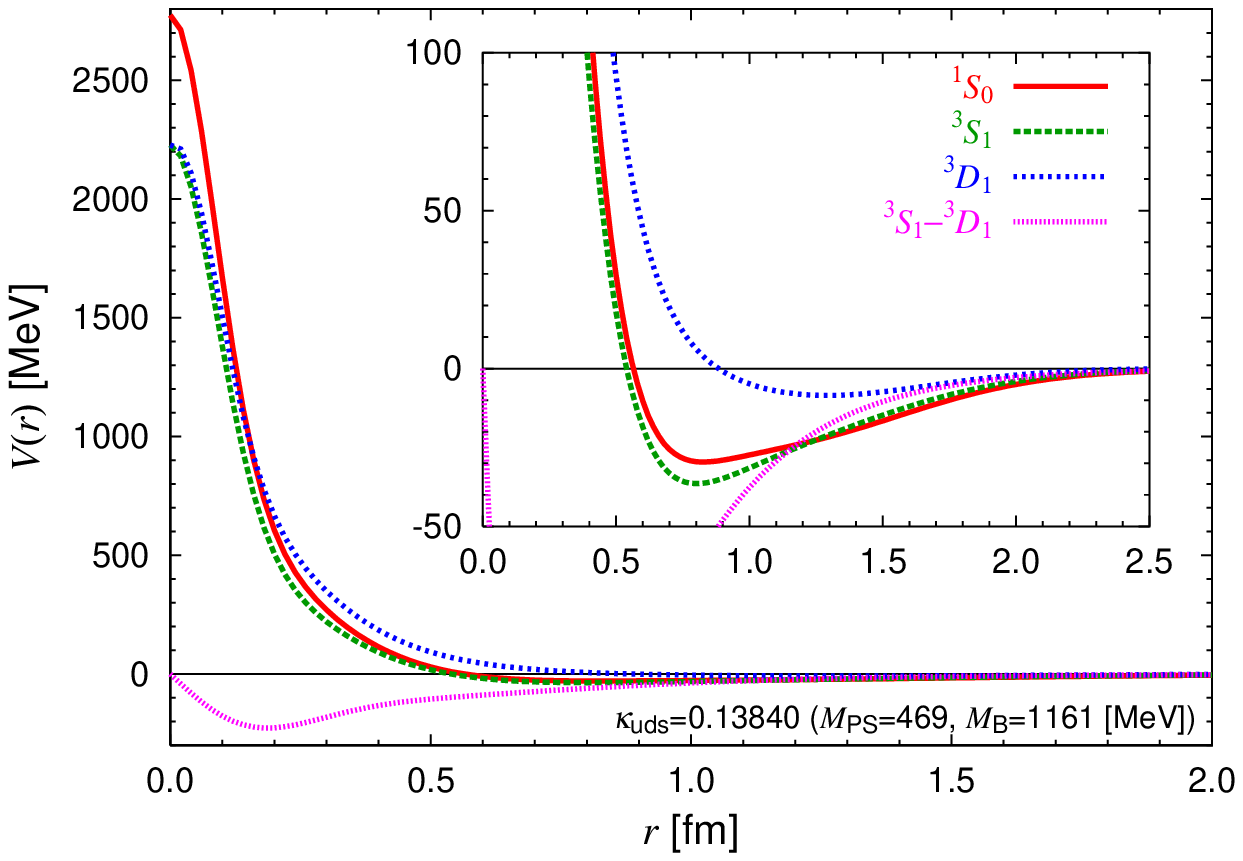}
\caption{Lattice QCD induced potentials of $NN$ interaction in the partial wave basis,
 extracted at a quark mass corresponding to $M_{\rm PS}=469$ MeV. Analytic functions fitted to data are plotted.}
\label{fig:pot_K13840}
\end{minipage}%
\hfill
\begin{minipage}[t]{0.475\columnwidth}
\centering
 \includegraphics[width=1.0\textwidth]{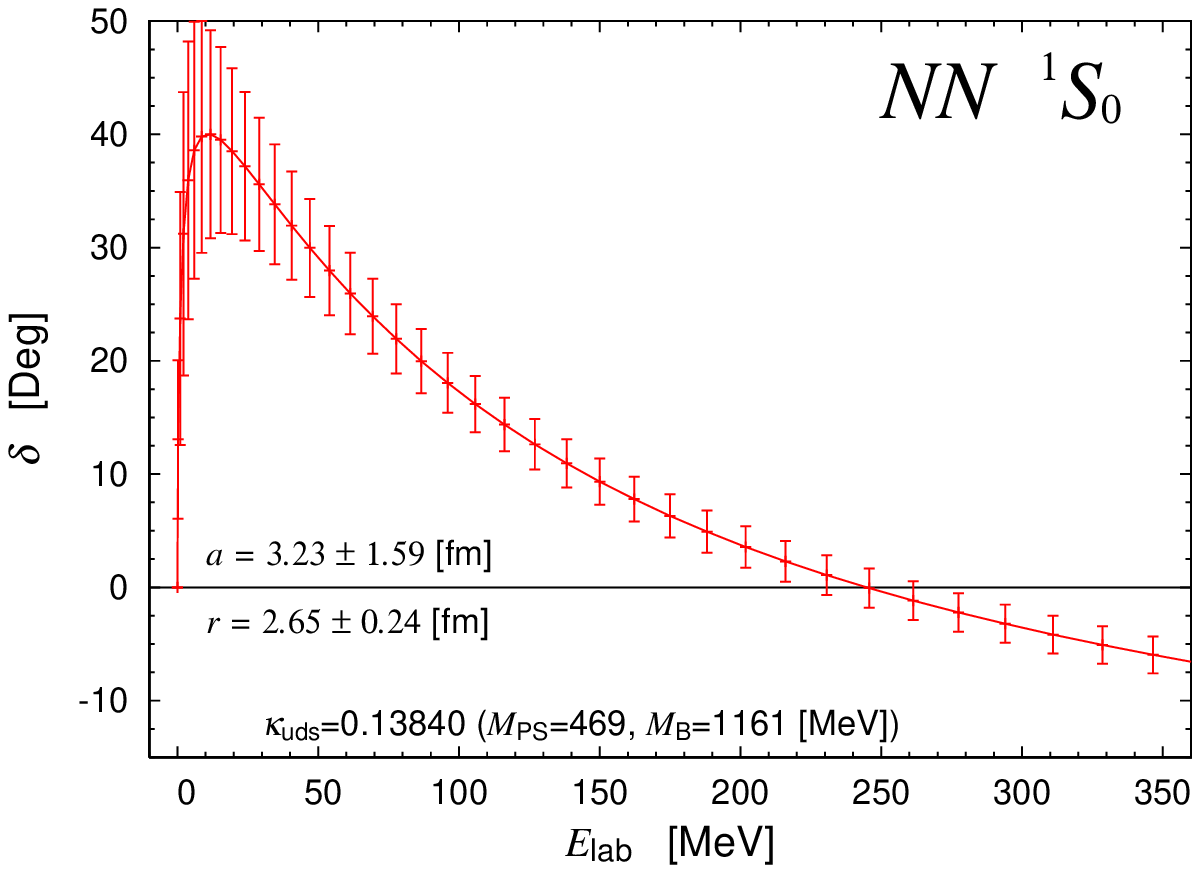}
\caption{Scattering phase shift of $NN$ $^1S_0$ partial wave obtained with a lattice QCD nuclear force
at a quark mass corresponding to $M_{\rm PS}=469$ MeV, as a function of the laboratory energy $E_{\rm lab}$.}
\label{fig:phase}
\end{minipage}
\end{figure} 

Fig.~\ref{fig:pot_K13840} shows potentials of $N\!N$ interaction in the partial wave basis, 
extracted from lattice QCD at a quark mass corresponding to $M_{\rm PS}=469$ MeV.
There, analytic functions fitted to data are plotted. For example, an analytic function  
\begin{equation}
 V(r) = b_1 e^{-b_2\,r^2}
      + b_3 e^{-b_4\,r^2}
      + b_5 \left( (1 - e^{-b_6\,r^2}) \frac{e^{-b_7\,r}}{r} \right)^2
\end{equation}
is used for the central potentials.
We use these analytic form of potentials in evaluating their matrix elements to study physical observables.
Fig.~\ref{fig:phase} shows phase shift of $N\!N$ scattering in the $^1S_0$ partial wave
obtained with the leading order potential $V(r)$, as a function of the laboratory energy $E_{\rm lab}$. 
The vertical bars contain only statistical error estimated in the Jackknife method.
There should be sizable systematic error in addition to statistical ones, 
especially at large laboratory energies due to the truncation of higher order terms in the derivative expansion.
Nevertheless, one can realize from this figure that the lattice QCD $NN$ potentials reproduce
the aspects of two-nucleon scattering observables very well. 

%% file: section_4.tex
\section{Helium nucleus from QCD}
\label{sec:he4}

\begin{figure}[t]
\begin{minipage}[t]{0.475\columnwidth}
\centering
\includegraphics[width=0.70\textwidth]{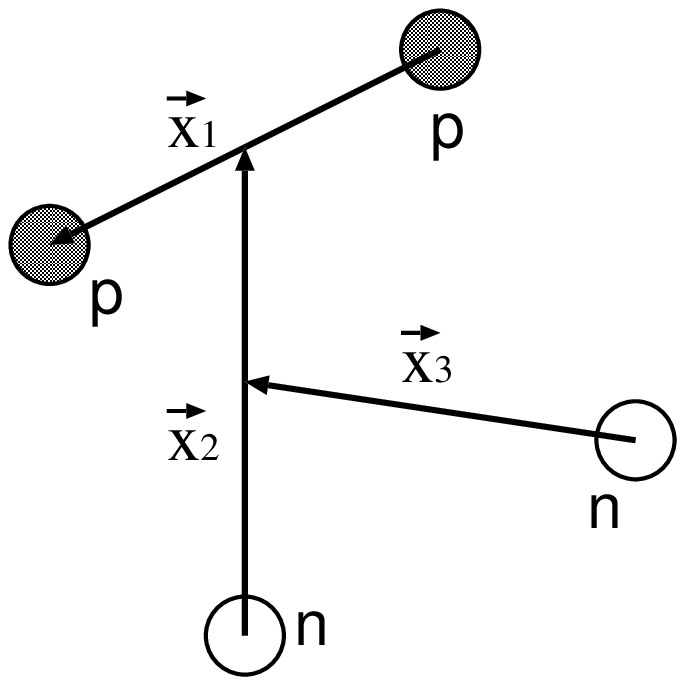}
\caption{The Jacobi coordinates for a four nucleon system, for example $^4\mbox{He}$.}
\label{fig:jacobi}
\end{minipage}%
\hfill
\begin{minipage}[t]{0.475\columnwidth}
\centering
\includegraphics[width=1.0\textwidth]{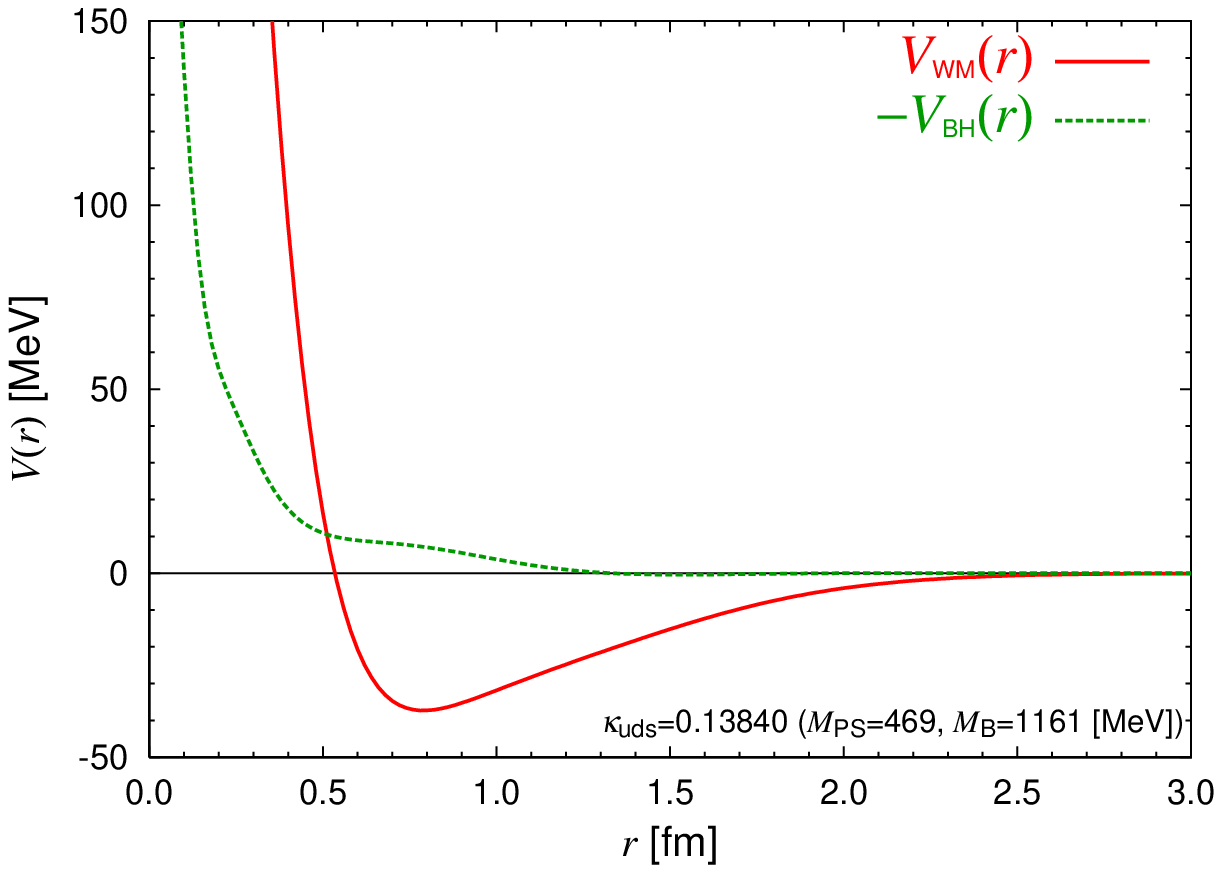}
\caption{Two terms which make up the central part of $NN$ potential in even parity partial wave,
determined with a lattice QCD nuclear force at a quark mass corresponding to $M_{\rm PS}=469$ MeV.}
\label{fig:pot_wmbh}
\end{minipage}
\end{figure}

In this section, we study few-nucleon systems using the lattice QCD induced $NN$ potentials.
We solve the Schr\"odinger equation given for example for $^4\mbox{He}$ case by
\begin{equation}
\left[ K + V \right] \, \Psi(\vec x_1,\vec x_2,\vec x_3) = E \, \Psi(\vec x_1,\vec x_2,\vec x_3)
\end{equation}
where $K$ ($V$) is the kinetic (potential) term of the Hamiltonian,
and $\{ \vec x_1, \vec x_2, \vec x_3 \}$ are the Jacobi coordinates shown in Fig.~\ref{fig:jacobi}. 
In this section, we deal with central potentials to simplify solving the equation.
We take into account the effect of the tensor force partially by means of the effective central potential.

In general, the central part of two nucleon potential $V_C(r)$ can be divided into
the Wigner $V_{\rm W}(r)$, Majorana $V_{\rm M}(r)$, Bartlett $V_{\rm B}(r)$, and Heisenberg $V_{\rm H}(r)$ as
\begin{equation}
V_C(r) = V_{\rm W}(r) + V_{\rm M}(r) P^{r} + V_{\rm B}(r) P^{\sigma} + V_{\rm H}(r) P^{r} P^{\sigma}
\end{equation}
where $P^{r}$ and $P^{\sigma}$ are permutation operators in space and spin-space respectively. 
For even parity partial waves, this decomposition is reduced to
\begin{equation}
V_C(r) = (V_{\rm W}(r) + V_{\rm M}(r)) + (V_{\rm B}(r) + V_{\rm H}(r)) P^{\sigma}
     \equiv V_{\rm WM}(r) + V_{\rm BH}(r) P^{\sigma}  ~.
\end{equation}
We determine these $V_{\rm WM}(r)$ and $V_{\rm BH}(r)$
by using data of lattice QCD $NN$ potential in $^1S_0$ and $^3S_1$ partial waves.
Fig.~\ref{fig:pot_wmbh} shows $V_{\rm WM}(r)$ and $V_{\rm BH}(r)$
determined with data at the present lightest quark mass corresponding to $M_{\rm PS}=469$ MeV.
There, we have used data of the effective central potential for $^3S_1$ partial wave,
in order to partially take into account the contribution from the tensor force.
We see that $V_{\rm BH}(r)$ is negative and much weaker than $V_{\rm WM}(r)$.

We do not have lattice QCD $NN$ potentials for odd parity partial waves.
Therefore, we consider two cases: the Wigner type force and the Serber type force.
In the Wigner type force, we set
\begin{equation}
 V_{\rm W}(r) = V_{\rm WM}(r),\quad V_{\rm M}(r) = 0,\quad V_{\rm B}(r) = V_{\rm BH}(r), \quad V_{\rm H}(r) = 0
\end{equation}
so that potential acting on odd parity partial waves is equal to one acting on even parity partial waves.
While, in the Server type force, we set
\begin{equation}
 V_{\rm W}(r) = \frac{V_{\rm WM}(r)}{2},\quad
 V_{\rm M}(r) = \frac{V_{\rm WM}(r)}{2},\quad
 V_{\rm B}(r) = \frac{V_{\rm BH}(r)}{2},\quad
 V_{\rm H}(r) = \frac{V_{\rm BH}(r)}{2}
\end{equation}
so that potential acting on odd parity partial waves is absent.
We consider these two extreme cases and compare results.
Since odd parity partial wave is known to be negligible in the S-shell nuclei, 
two results will almost coincide, and our approximation for unknown odd parity potential should be reasonable,
for at least three- and four-nucleon systems.

In order to solve the Schr\"odinger equation of few-body systems,
we use the stochastic variational method~\cite{Varga:1995dm}
where the correlated Gaussian bases are used to expand the wave function $\Psi$.
The correlated Gaussian basis, for total angular momentum $L=0$, is given by
\begin{equation}
f_A(\vec x_1,\vec x_2,\vec x_3) = \exp\left[ - \frac12 X\cdot A X^t \right]
\end{equation}
where $X=(\vec x_1,\vec x_2,\vec x_3)$ and $A$ is a symmetric and positive definite $3 \times 3$ matrix.
By generating the matrix $A$ randomly, many functions $f_A$ are examined. 
Then, the most efficient one for the state of interest is added to the basis set. 
This is what is called competitive selection. The number of basis functions gradually increases but remains small, 
because energies converge rapidly since important basis functions are selected.
This means that we do not need to prepare a huge basis set
and not need to diagonalize a huge Hamiltonian matrix from the beginning. 
Therefore, it is easy to solve the Schr\"odinger equation of few-body systems in this method.

\begin{figure}[t]
\begin{minipage}[t]{0.475\columnwidth}
\centering
\includegraphics[width=1.0\textwidth]{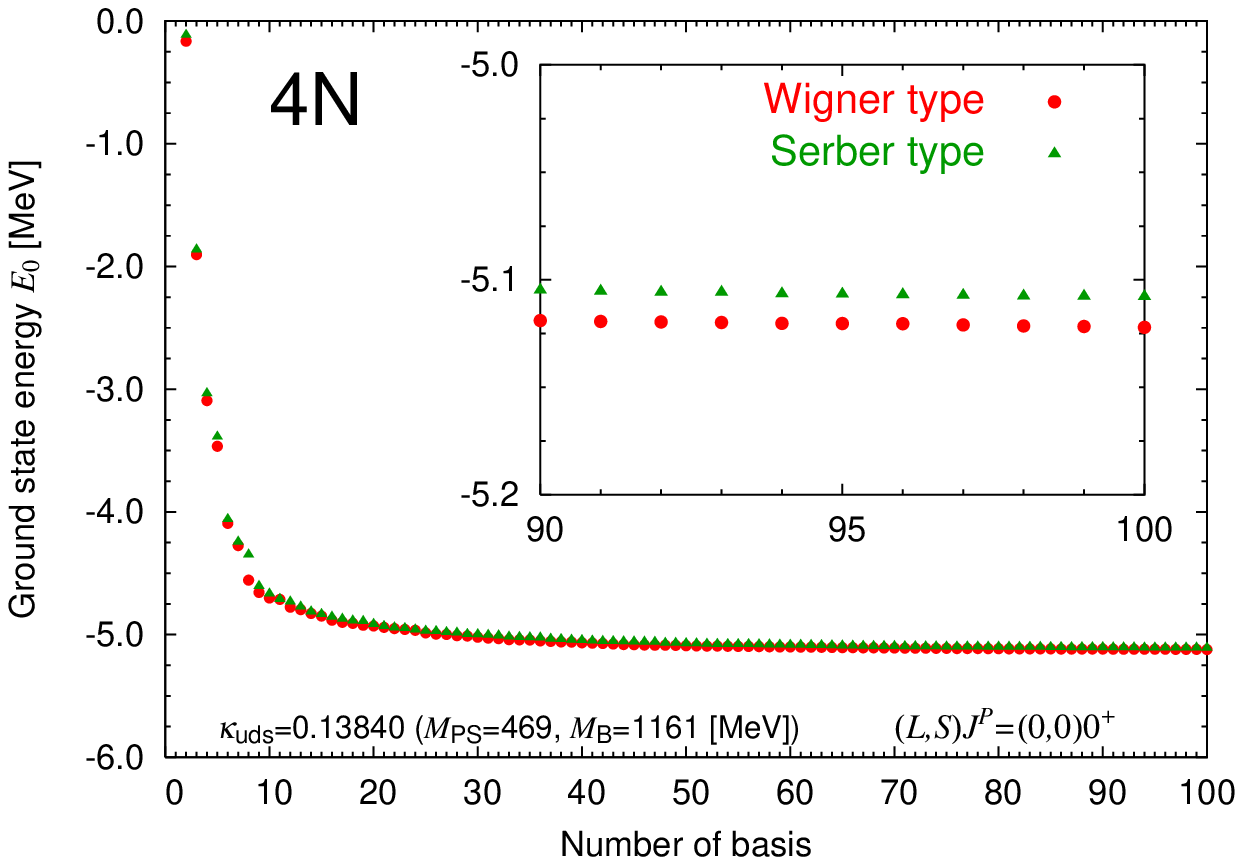}
\caption{Ground state energy of four-nucleon system 
obtained with a lattice QCD $NN$ potentials at $M_{\rm PS}=469$ MeV, as a function of number of basis funnctions.
Two types of two-nucleon interactions are compared. See the text for details.}
\label{fig:alpha}
\end{minipage}%
\hfill
\begin{minipage}[t]{0.475\columnwidth}
\centering
\includegraphics[width=1.0\textwidth]{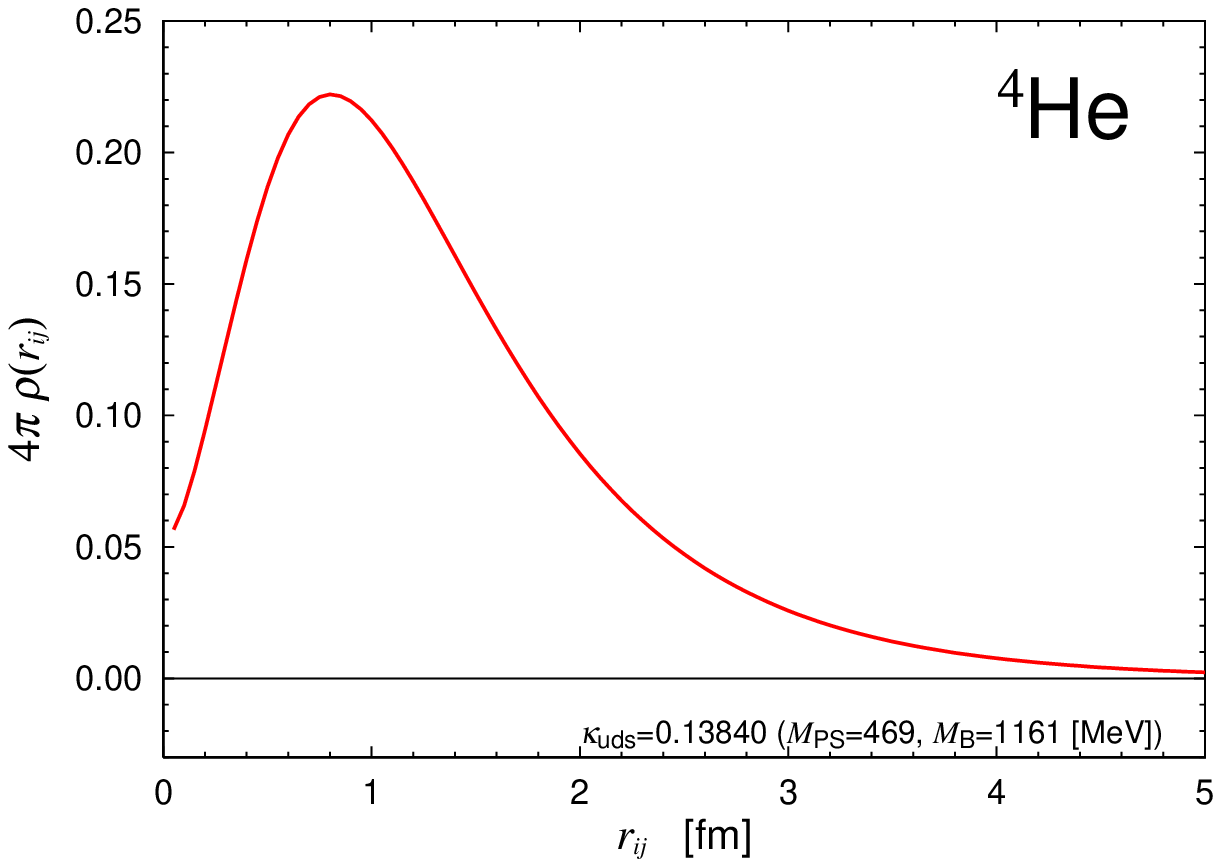}
\caption{Correlation function of two-nucleon in the $^4$He ground state,
obtained with a lattice QCD $NN$ potentials at $M_{\rm PS}=469$ MeV.}
\label{fig:he4_rho_r12}
\end{minipage}
\end{figure}

Fig.~\ref{fig:alpha} shows the lowest energy eigenvalue of four-nucleon system
with isospin $I=0$ and $(L,S)J^P=(0,0)0^+$ configuration, 
obtained with the lattice QCD $NN$ potentials at $M_{\rm PS}=469$ MeV, as a function of number of basis functions. 
One see that energy converges rapidly,
and that result with the Wigner-type force and that with the Serber-type force almost agree
as expected so that the effect of odd parity potential is negligible.
This bound state corresponds to the ground state of $^4$He nucleus.
Namely, we find a stable $^4$He nucleus in a QCD world with this quark mass.
We observe that the binding energy of $^4$He at this quark mass is about 5.1 MeV.
Note that the three-nucleon and four-nucleon forces are not considered and may change the binding energy a little.
Obtained binding energy is much smaller than the experimental value of $28.295$ MeV.
This discrepancy is primarily due to the unphysical quark mass in our study.
Fig.~\ref{fig:he4_rho_r12} shows correlation function of two-nucleon in the $^4$He ground state at this quark mass.
We can see effect of the repulsive core at short distance.

An indication of very shallow four-nucleon bound state is seen
for the heaviest quark mass and the second heaviest quark mass
corresponding to pseudo-scalar meson mass $M_{\rm PS}=1171$ MeV and $M_{\rm PS}=1015$ MeV, respectively.
Since the obtained binding energy is tiny, we do not draw a conclusion about these signals.

We do not find any four-nucleon bound state for the second lightest and middle quark mass
corresponding to $M_{\rm PS}=672$ MeV and $M_{\rm PS}=837$ MeV, respectively.
And, we do not find any two-nucleon and three-nucleon bound state for all the five values of quark mass
corresponding to a range of pseudo-scalar meson mass from $M_{\rm PS}=469$ MeV to $M_{\rm PS}=1171$ MeV.
From these results, we conclude that light nuclei are diffcult to bind at large quark mass.
This conclusion is in contrast to results of lattice QCD studies by other groups,
where strongly bound light nuclei \ie two-nucleon, three-nucleon, and four-nucleon bound states are reported.
Since those groups use the direct method in which binding energy of multi-nucleon system is directly extracted
from lattice QCD temporal correlation function, we suspect that their results suffer from the plateau crisis.

%% file: section_5.tex
\section{Medium-heavy nuclei from QCD}
\label{sec:medium}

\begin{figure}[t]
\centering
\includegraphics[width=0.95\textwidth]{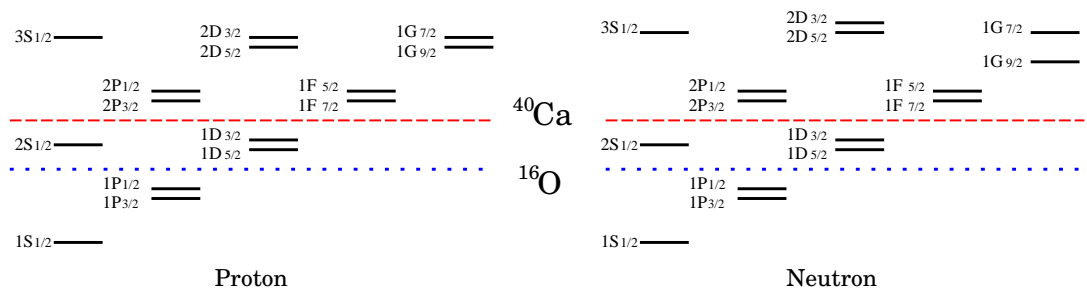}
\caption{A schematic diagram of single particle (quasi nucleon) levels in nuclei in the mean field picture.
Levels below blue dotted (red dashed) line are occupied completely in $^{16}$O \ ($^{40}$Ca) nucleus.}
\label{fig:meanfield}
\end{figure} 

In this section, we study medium heavy nuclei starting from QCD in the HAL QCD approach.
Unfortunately, we cannot solve medium heavy nuclei exactly as we did in the previous section.
So, let us begin with the mean field picture of (medium) heavy nuclei. 
Fig. \ref{fig:meanfield} shows a schematic diagram of single particle (quasi nucleon) levels in nuclei.
For example, levels below the blue dotted line are occupied completely in $^{16}$O nucleus,
while ones below the red dashed line are filled completely in $^{40}$Ca nucleus.
This single particle picture was proven to be very useful for (medium) heavy nuclei.
In fact, the nuclear shell model has achieved many successes.
By the way, $^{16}$O and $^{40}$Ca are called doubly closed or doubly magic nuclei,
since there the major shells (set of levels with almost degenerate energy),
are occupied completely or not occupied at all for both proton and neutron.

The above independent particle nature of nucleonic system was explained microscopically
based on a two-nucleon interaction in free space, by the Brueckner theory~\cite{Brueckner:1958zza}.
Consequently, the Brueckner-Hartree-Fock (BHF) theory became a standard framework
to obtain (medium) heavy nuclei based on a bare interaction.
After that, nuclear theory continued to develop, and today we have several sophisticated theories for (medium) heavy nuclei beyond the BHF theory.
For example, recent studies show that the coupled-cluster theory~\cite{Hagen:2007hi}, 
the unitary-model-operator approach~\cite{Fujii:2009bf}, and the self-consistent-Green's function method~\cite{Dickhoff:2004xx}
are powerful for these nuclei and even better than the BHF theory.
Moreover, ab initio calculations are carried out successfully for nuclei around $^{12}$C,
in the Green's function Monte Carlo method~\cite{Pieper:2001mp} and the no-core shell model~\cite{Navratil:2000ww,Shimizu:2012mv},
although exact application to heavier nuclei seems difficult at this moment.

Since this study is our first attempt to attack medium heavy nuclei starting from QCD,
we employ the traditional BHF theory in this paper.
The BHF theory is simple but quantitative enough to grasp the essential part of physics
so that this study is good starting point before making precise calculations using modern sophisticated theories.
Below, we try to obtain properties of $^{16}$O nucleus and $^{40}$Ca nucleus in the BHF theory.
We choose these two nuclei because they are doubly magic nucleus,
whose ground state is safely assumed as isospin symmetric, spin saturated, and spherically symmetric,
and hence our BHF calculation become easy.

In the BHF theory, $G$ matrix which describes scattering of two quasi nucleons, 
is a important ingredient and obtained by solving the integral Bethe-Goldstone equation 
\begin{equation}
 G(\omega)_{ij,kl} \,=\, V_{ij,kl} \,+\,
 \frac12 \, \sum_{m,n}^{>e_F} \, \frac{V_{ij,mn}\,G(\omega)_{mn,kl}}{\omega - e_m - e_n + i\epsilon}
\label{eqn:gmat}
\end{equation}
where indices $i$ to $n$ stand for a single-particle energy-eigenstates
and $V$ is a $N\!N$ interaction potential and the intermediate sum runs over excluding occupied states of the nucleus.
With this $G$ matrix, the single-particle potential $U$ is given by
\begin{equation}
 U_{ab} = \sum_{c,d} G(\tilde{\omega})_{ac,bd}~\rho_{dc}
\label{eqn:sppotU}
\end{equation}
where indices $a,b,c,d$ correspond to a basis-functions and $\rho$ is the density matrix in this basis,
which is given with the wave function of energy-eigenstate $\Psi^{i}$ by
\begin{equation}
 \rho_{ab} = \sum_{i}^{\rm occ} \Psi^{i}_a \Psi^{i*}_b
\label{eqn:rhomat}
\end{equation}
where the sum runs over occupied states of the nucleus.
However, the energy-eigenstates are obtained as a solution of
the Hartree-Fock equation involving the potential $U$
\begin{equation}
 \left[ K + U \right] \Psi^{i} = e_i \Psi^{i}
\label{eqn:scheq}
\end{equation}
where $K$ is the kinetic energy operator of nucleon.
Because these equations are highly coupled, 
self-consistent $G$, $U$, $\rho$, $\Psi^{i}$ and $e_i$ are determined by an iteration procedure.
Finally, the Hartree-Fock ground state energy of the nucleus $E_0$
is obtained with the self-consistent $U$ and $\rho$ by
\begin{equation}
 E_{0} = \sum_{a,b} \left[ K_{a b} + \frac12 U_{a b} \right] \rho_{b a} - K_{\rm cm}
\label{eqn:gsene}
\end{equation}
where $K_{\rm cm}$ is the kinetic energy corresponding to the spurious center-of-mass motion
in the potential rest frame which is included in $K$ in the first term.

We carry out the above BHF calculation by using the lattice QCD nucleon mass $M_B$
and the lattice QCD induced two-nucleon potentials $V(r)$ shown in the section~\ref{sec:setup}. 
Due to the limitation for the lattice QCD $N\!N$ potentials available at present,
we include nuclear force only in $^1S_0$, $^3S_1$ and $^3D_1$ channels.
We ignore the Coulomb force between protons for simplicity.
For details of the BHF calculation, we essentially follow refs.~\cite{Daveis,Sauer,Schmid:1991xp}. 
Namely, we use the harmonic-oscillator wave functions
\begin{equation}
 R_{nl}(r)= \sqrt{\frac{2 n!}{\Gamma(n+l+\frac32)}}\left(\frac{r}{b}\right)^l
            e^{-\frac12 \frac{r^2}{b^2}}
            \sum_{m=0}^{n} C^{n+l+\frac12}_{n-m} \frac{\left(-{r^2}/{b^2}\right)^m }{m!}
\label{eqn:Rnl}
\end{equation}
for the basis-functions,
and solve eq.(\ref{eqn:gmat}) by separating the relative and center-of-mass coordinates by using the Talmi-Moshinsky coefficient.
We use the angle-averaged Pauli exclusion operator $Q$, and adopt the so-called $Q/(\omega - QKQ)Q$ choice.
We begin with the harmonic-oscillator $Q$, then use the self-consistent $Q$ in the last few iterations.
For $\tilde{\omega}$ in eq.(\ref{eqn:sppotU}), we use the standard prescription used in ref.~\cite{Daveis}.
For the center-of-mass correction in eq.(\ref{eqn:gsene}), we use the estimate $K_{\rm cm} = \frac{3}{4}\hbar \omega$
with $\omega$ being the a harmonic-oscillator frequency, which reproduces the root-mean-square radius
of the point matter distribution obtained in the BHF calculation~\cite{Lipkin:1958zz}.

\begin{figure}[t]
\includegraphics[width=0.45\textwidth]{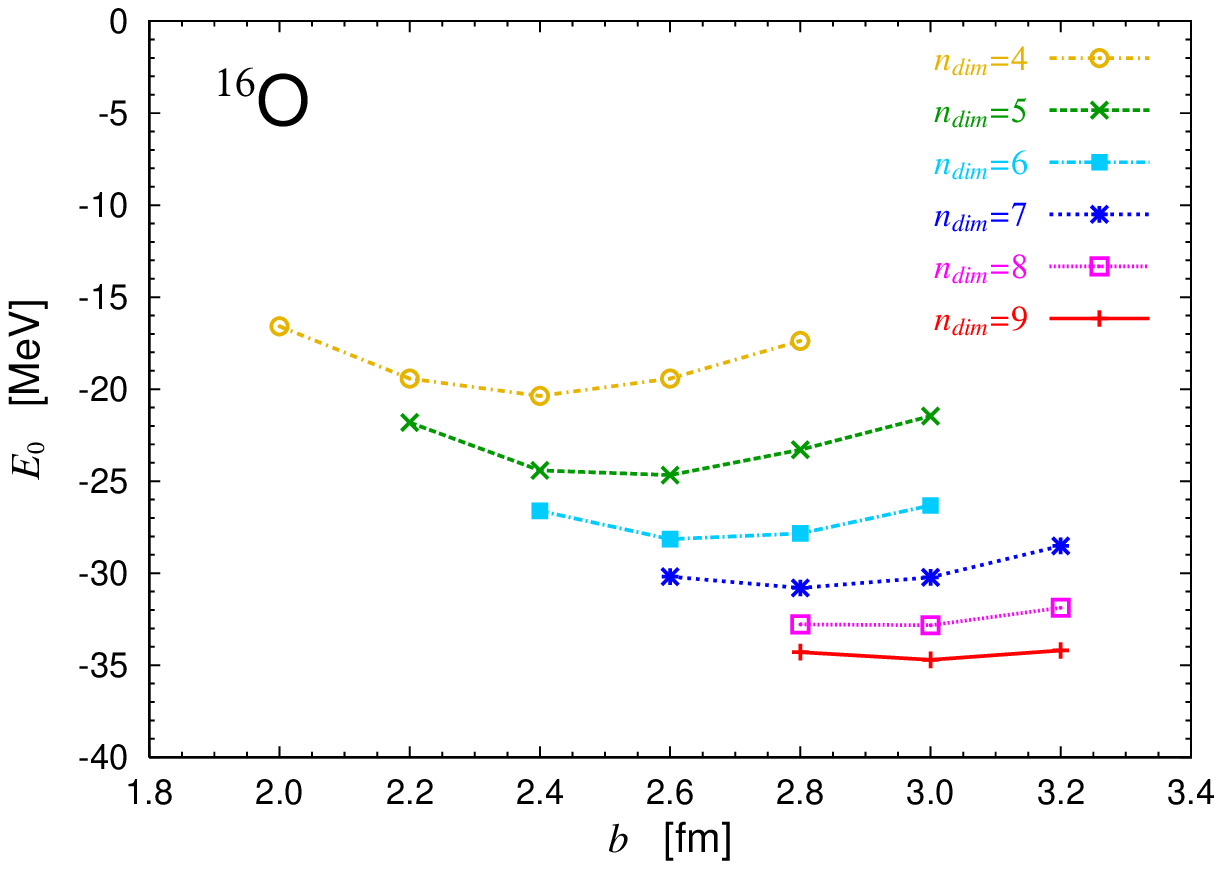}~~
\includegraphics[width=0.45\textwidth]{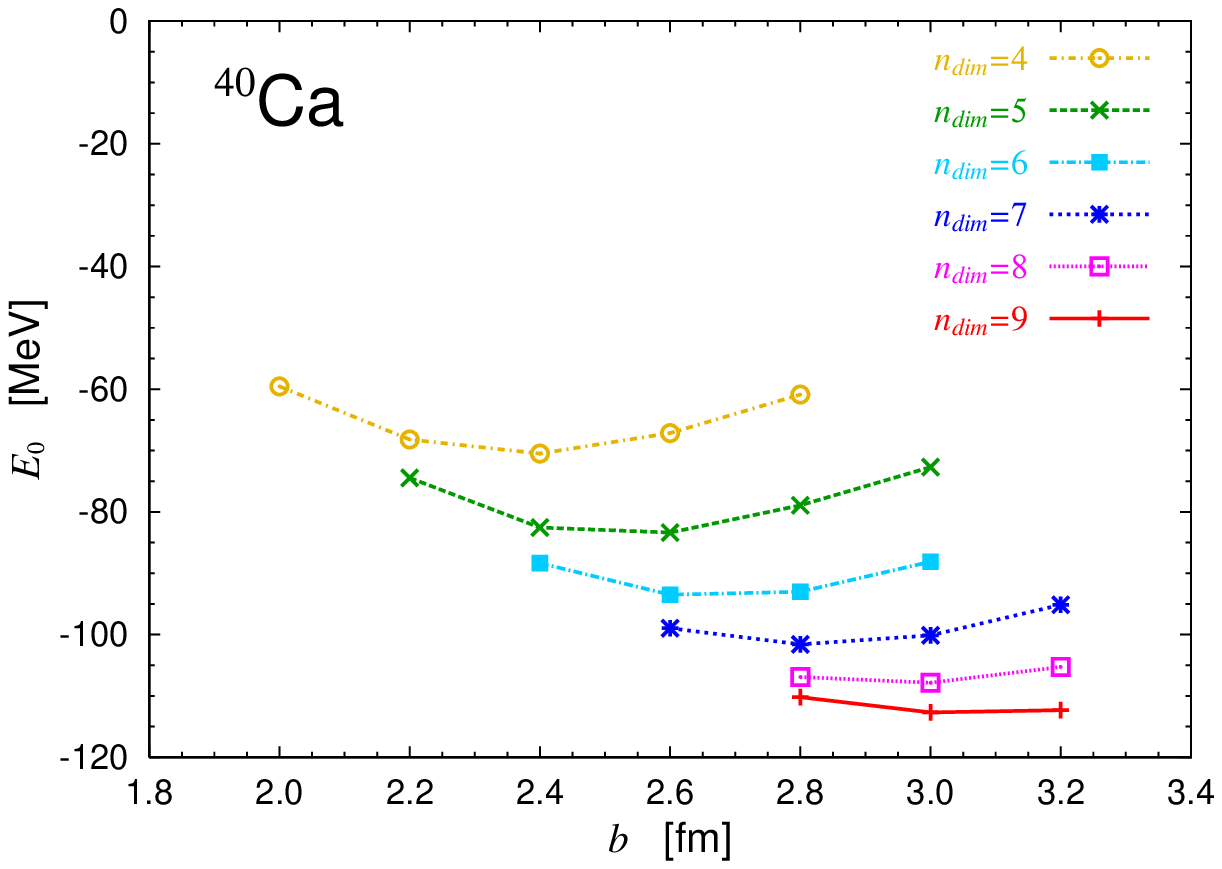}
\caption{Ground state energy of $^{16}$O and $^{40}$Ca obtained in our BHF calculation
with a lattice QCD nuclear force at a quark mass corresponding to $M_{\rm PS}=469$ MeV,
as a function of the length parameter $b$ of the harmonic-oscillator basis-function,
at several $n_{\rm dim}$ the number of the basis for each angular momentum.}
\label{fig:nbdepend}
\end{figure} 

To begin with, we investigate whether the medium-heavy nuclei exist or not,
and search optimal harmonic-oscillator basis-functions and number of basis-functions needed to solve the nuclei.
Fig.~\ref{fig:nbdepend} shows ground state energy $E_0$ of $^{16}$O and $^{40}$Ca obtained in our BHF calculation 
at our lightest quark mass corresponding to $M_{\rm PS}=469$ MeV,
as a function of the length parameter $b$ of the harmonic-oscillator basis-functions,
for the increasing size of basis $n_{dim}$ for each angular momentum $l$. 
We see that the energy depends on the parameter $b$ and its convergence in $n_{\rm dim}$ is slow.
However, we see also that these energies are definitely sufficiently negative. 
In addition, the resulting binding energies are larger than four or ten times the $^4$He biding energy in the previous section.
Consequently, we conclude that there are stable $^{16}$O and $^{40}$Ca nuclei at this quark mass.
This is the first-ever finding of medium-heavy nuclei in lattice QCD~\cite{Inoue:2014ipa}.

On the other hand, we do not obtain any negative $E_0$ for both $^{16}$O and $^{40}$Ca
in our BHF calculation at the other four values of quark mass.
This at least means that there is no tightly-bound nucleus in QCD at these values of quark mass.
Therefore, in the following, we consider only our lightest quark mass case. 
Note that nucleon mass is 1161 MeV and pion mass is 469 MeV in this case.
Because increasing $n_{\rm dim}$ more is tough for our computer system, we adopt $n_{dim}=9$ in this paper.
We use $b=3.0$ fm for both $^{16}$O and $^{40}$Ca as suggested by the figure.

\begin{table}[t]
\caption{Single particle levels, total energy, and mean radius of $^{16}$O and $^{40}$Ca
obtained in BHF calculation with a lattice QCD nuclear force at a quark mass corresponding to $M_{\rm PS}=469$ MeV.}
\label{tbl:structure}
\medskip
\centering
\begin{tabular}{c|cccc|cc|c}
\hline \hline
       & \multicolumn{4}{c|}{Single particle level ~[MeV]}
       & \multicolumn{2}{|c|}{Total energy ~[MeV]} & \,Radius ~[fm]\, \\
       & ${1S}$  & ${1P}$  & ${2S}$  & ${1D}$  & $E_0$  & $E_{0}/A$ & $\sqrt{\langle r^2 \rangle}$  \\
\hline 
 $^{16}$O~  & \,$-34.1$\,  & \,$-13.2$\,  &              &           &   ~$-32.8$    & $-2.05$     &   $2.44$  \\
 $^{40}$Ca  & \,$-55.7$\,  & \,$-34.3$\,  & \,$-13.4$\,  & \,$-14.1$ & \,$-107.9$\,  & \,$-2.70$\, & \,$2.89$\,\\
 \hline \hline
\end{tabular}
\end{table}

\begin{figure}[t]
\begin{minipage}[t]{0.475\columnwidth}
\centering
\includegraphics[width=1.0\textwidth]{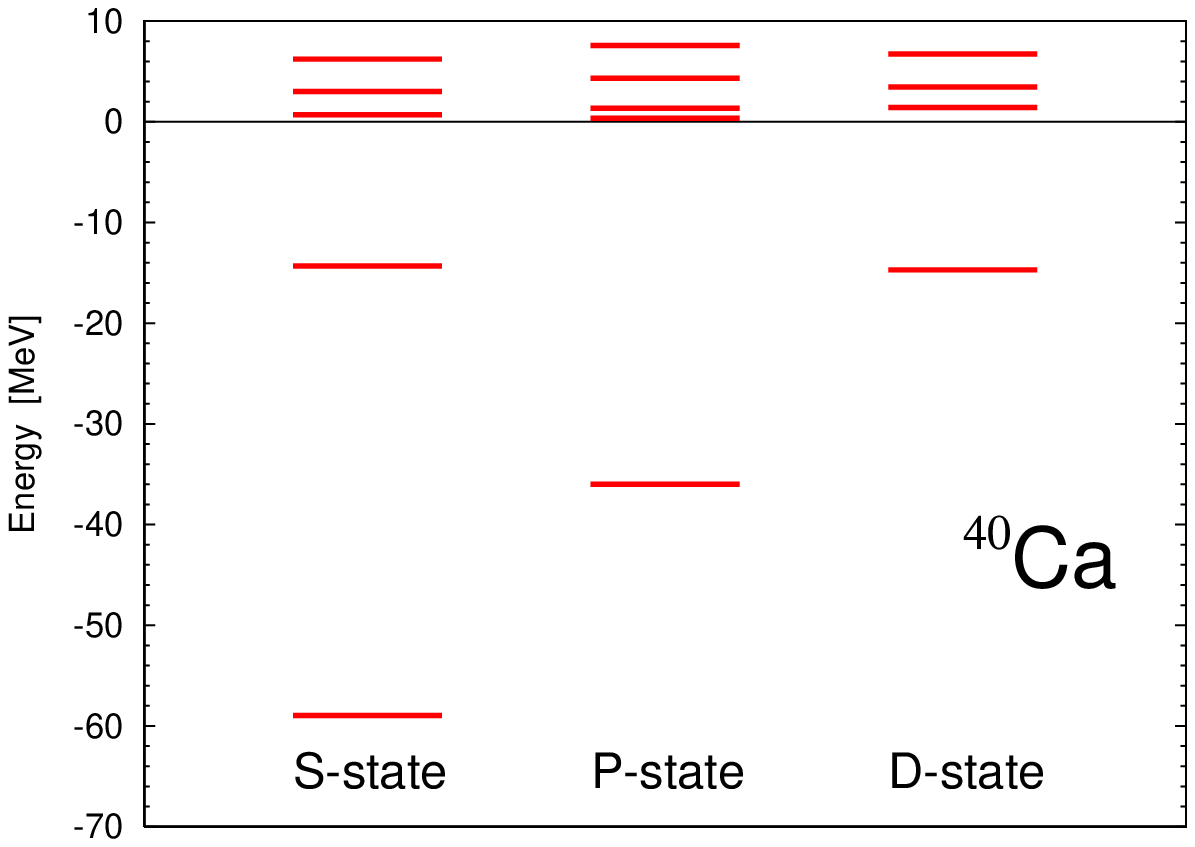}
\caption{Single particle levels in the $^{40}$Ca at a quark mass of $M_{\rm PS}=$ 469 MeV.
The positive energy continuum appears as discrete levels because of the finite number of basis-functions.}
\label{fig:levels}
\end{minipage}%
\hfill
\begin{minipage}[t]{0.475\columnwidth}
\centering
\includegraphics[width=1.0\textwidth]{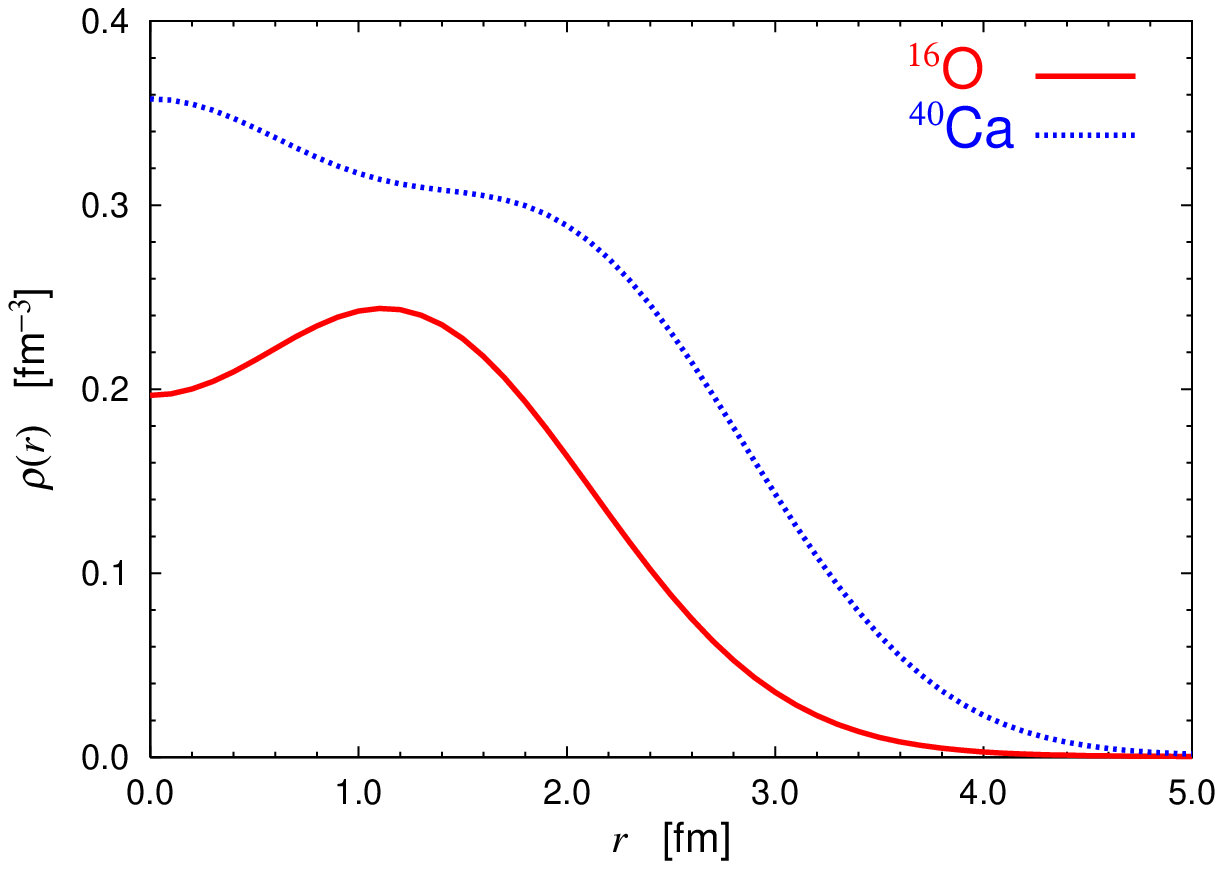}
\caption{Point-nucleon density distribution inside $^{16}$O and $^{40}$Ca 
at a quark mass of $M_{\rm PS}=$ 469 MeV, as a function of distance from the center.}
\label{fig:density}
\end{minipage}
\end{figure}

Table~\ref{tbl:structure} shows single particle levels, total energy, and root-mean-square radius
of the ground state of the two nuclei, obtained at the lightest quark mass.
The single particle levels of $^{40}$Ca are shown in Fig.~\ref{fig:levels}.
There, we can see regular shell structure clearly.
These levels are already in good agreement with experimental data
which can be found in for example Table 9 of ref.~\cite{Negele:1970jv}.
However, this agreement might be accidental because
we have used the unphysical value of nucleon mass and several approximations in our BHF calculation.

Obtained total energy of the ground state is $-32.8$ MeV for $^{16}$O and $-107.9$ MeV for $^{40}$Ca, 
whose breakdown are given by
\begin{eqnarray}
 ^{16}\mbox{O}:  && E_0 ~=~ 247.9 ~-~ 271.2 ~-~ 9.5 ~=~ ~\,-32.8 ~~ \mbox{[MeV]}\\
 ^{40}\mbox{Ca}: && E_0 ~=~ 772.6 ~-~ 871.4 ~-~ 9.0 ~=~   -107.9 ~~ \mbox{[MeV]}
\label{eqn:break}
\end{eqnarray}
where the first and second number is kinetic and potential energy, respectively,
and the last number is the center-of-mass correction $K_{\rm cm}$ estimated.
As usual, total energy is obtained as a result of very subtle cancellation between kinetic energy and potential energy.
Therefore, we should probably take the above $E_0$ only qualitative since they are obtained with several approximations.
Of course, obtained $E_0$ are much smaller than the experimental data
which is $-127.62$ MeV for $^{16}$O and $-342.05$ MeV for $^{40}$Ca~\cite{Audi:1993zb}.
Again, this discrepancy is primarily due to the unphysical quark mass in our study.
Recall that we have not used any phenomenological input for nucleon mass and nucleon interaction, but have used only QCD.

Fig.~\ref{fig:density} shows nucleon density distribution inside the nuclei as a function of the distance from the center. 
The root-mean-square radii of the distribution are given in Table~\ref{tbl:structure}.
These are calculated with a point-nucleon and without taking the center-of-mass correction.
In the figure, we can see a bump and dent at small distance, which are effects of the shell structure.
These effects are observed in experimental charge distribution which can be found in \eg ref.~\cite{Sick:1970}.
Contrary to large discrepancies of $E_0$ from experimental data,
obtained radii are more of less in agreement with experimental charge radius, 2.73 fm for $^{16}$O and 3.48 fm for $^{40}$Ca.
Probably, this agreement is due to a cancellation between the weaker attraction in our potential and larger nucleon mass.
In summary, we have seen that the HAL QCD method combined with the many-body theory BHF
produces a reasonable structure of medium-heavy nuclei without using phenomenological input at all.

%% file: section_6.tex
\section{Nuclear matter equation of state from QCD}
\label{sec:matter}

In this section, we investigate equation of state (EoS) of nuclear matter starting from QCD in the HAL QCD approach.
Nuclear matter is a uniform matter consists of infinite number of nucleons.
Equation of state is an equation which gives the relation between energy and pressure of matter.
We need to obtain energy of the ground state of interacting infinite nucleon system.
One successful approach is the Brueckner-Bethe-Goldstone (BBG) expansion,
where perturbative expansion is rearranged in terms of the $G$ matrix,
and terms are ordered according to number of independent hole-lines appearing in its diagrammatic representation~\cite{Brueckner:1958zza}. 
The lowest-order two-hole-line approximation is called the Brueckner-Hartree-Fock (BHF) framework, 
which is nothing but the one adopted in the previous section.
We adopt the BHF framework in this section again.

In the BHF theory, $G$ matrix describing the scattering of two quasi nucleons is important.
It is depicted diagrammatically by a sum of ladder diagrams representing repeated action of the bare $N\!N$ interaction $V$, 
and obtained by solving the Bethe-Goldstone equation
\begin{equation}
   \langle k_1 k_2 |G(\omega)|  k_3 k_4 \rangle
 = \langle k_1 k_2|V|k_3 k_4 \rangle \nonumber 
 + \sum_{k_5, k_6}
    \frac{\langle k_1 k_2|V|k_5 k_6 \rangle \,Q(k_5,k_6) \, 
         \langle k_5 k_6 | G(\omega) |  k_3 k_4 \rangle}
         {\omega - e(k_5) - e(k_6)}
\label{eqn:gmateq}
\end{equation}
where $Q(k,k')=\theta(k-k_F)\theta(k'-k_F)$ is the Pauli exclusion operator
preventing two nucleons from scattering into the occupied states of matter \ie the Fermi sea, and $k_F$ being the Fermi momentum. 
The single particle spectrum, for nucleon mass $M_N$,
\begin{equation}
 e(k) = \frac{k^2}{2M_N} + U(k)
\end{equation}
contains a single particle potential $U(k)$, 
which is crucially important for faster convergence of the BBG expansion.
This potential is determined from Brueckner's consistency condition 
\begin{equation}
 U(k) = \sum_{k' \le k_F} \mbox{Re} \langle k k'| G(e(k)+ e(k')) |k k'\rangle_A 
\label{eqn:self}
\end{equation}
with $|k k'\rangle_A = |k k'\rangle - |k' k \rangle$. 
Because these equations are highly coupled, self-consistent $G$ and $U$ are determined by an iteration procedure.
Finally, total energy $E_0$ of the ground state of nuclear matter at zero temperature, is obtained with the self-consistent $G$ and $U$ by
\begin{equation}
 E_0 = \sum_{k}^{k_F} \frac{k^2}{2 M_N} 
     + \frac{1}{2} \sum_{k,k'}^{k_F} \mbox{Re} \langle k k'| G(e(k)+ e(k')) |k k' \rangle_A 
\end{equation}
where spin and isospin indices of the nucleons are included in the label $k$ to simplify the notation. 

We carry out the above BHF calculation by using the lattice QCD nucleon mass $M_B$
and the lattice QCD induced two-nucleon potentials $V(r)$ found in section~\ref{sec:setup}. 
We use the angle averaged $Q$-operator and decompose $G$-matrix in partial waves.
Again, we truncate the decomposition keeping $^1S_0$, $^3S_1$, and $^3D_1$ partial waves,
because of the limitations of our lattice QCD $NN$ potential.
We use the so called continuous choice of $U(k)$,
and the parabolic approximation of it in order to put it into the Bethe-Goldstone equation (\ref{eqn:gmateq}).

\begin{figure}[t]
\centering
\includegraphics[width=0.375\textwidth]{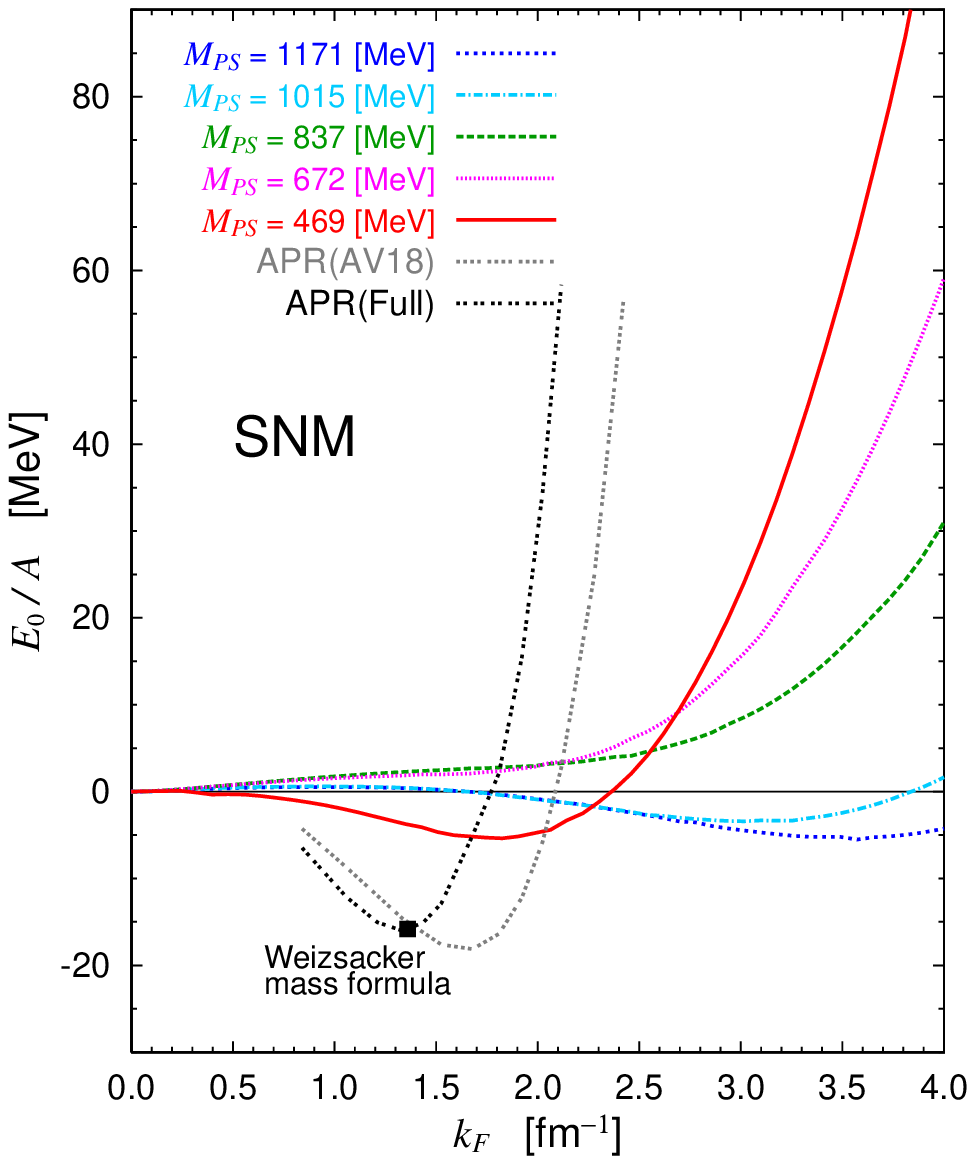}
\qquad
\includegraphics[width=0.375\textwidth]{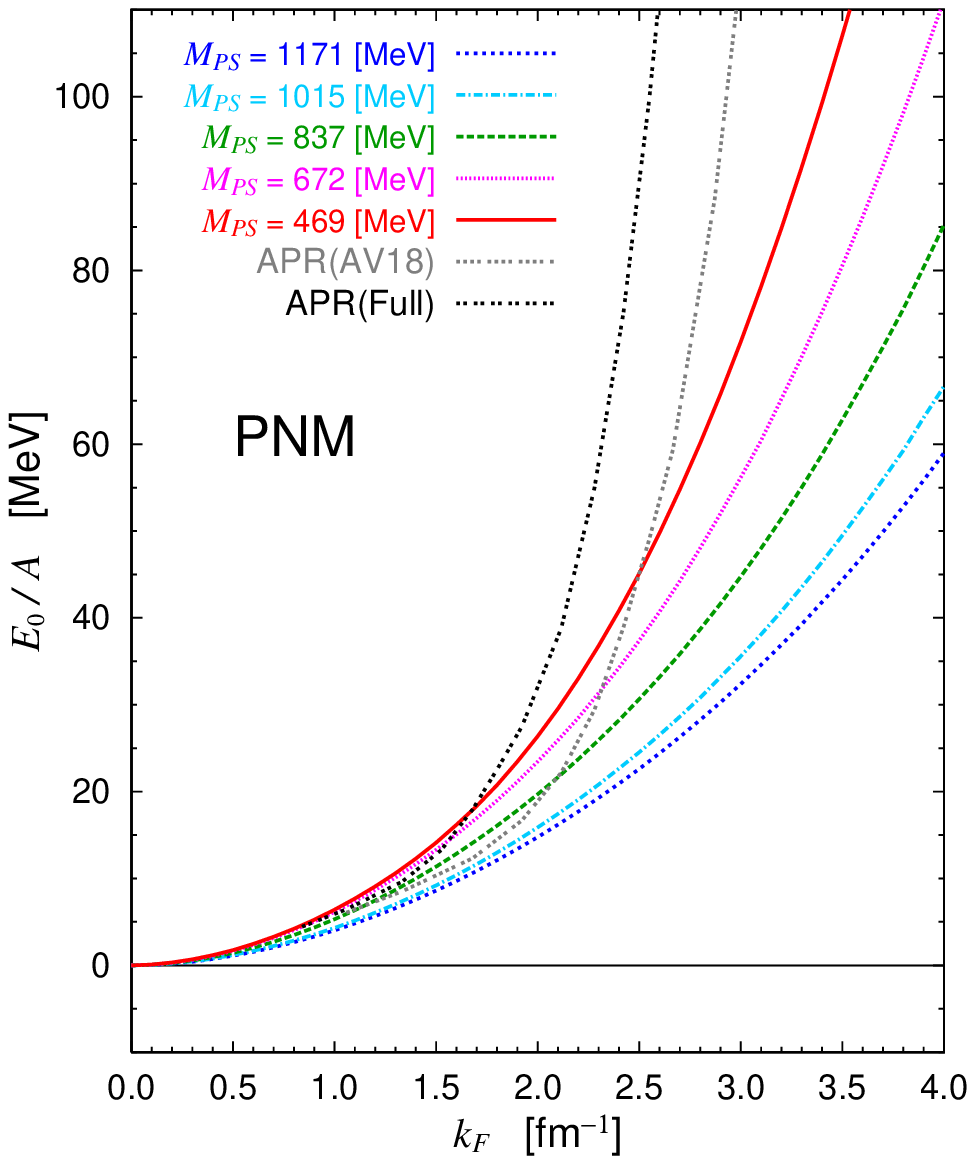}
\caption{Ground state energy per nucleon $E_0/A$ for symmetric nuclear matter in the left panel
and pure neutron matter in the right panel, as a function of the Fermi momentum $k_F$.  
The empirical saturation point is also indicated in the left panel.
The curves labeled APR are taken from ref.~\cite{Akmal:1998cf}}
\label{fig:eos}
\end{figure} 

Fig.~\ref{fig:eos}, in the left panel, shows the obtained ground state energy per nucleon $E_{0}/A$ 
for symmetric nuclear matter (SNM) as a function of the Fermi momentum $k_F$.
These curves are equivalent to the equation of state,
because the pressure of matter at a density is determined by a slope of the curve at that density. 
The most important feature of SNM is saturation in which both the binding energy per nucleon
and the nucleon density are constant independent on the number of nucleons $A$. 
The empirical saturation point, suggested from the Weizs\"acker mass formula and nuclear binding energy data,
is around $k_F = 1.36$ {fm}$^{-1}$ and $E_{0}/A = -15.7$ MeV, which is indicated in the figure.
In addition, the EoS reported in ref.~\cite{Akmal:1998cf} is shown with label APR for a reference.
The APR EoS is obtained in the Fermi-Hypernetted-chain variational calculation
with the physical nucleon mass, the modern phenomenological $N\!N$ potential Argonne V18, and a model $N\!N\!N$ force adjusted.
The APR EoS is often regarded as phenomenological in the literature.

We can see, in the left panel of Fig.~\ref{fig:eos}, 
that the SNM EoS obtained from QCD at the lightest quark mass corresponding to $M_{\rm PS}=469$ MeV clearly shows saturation.
This is the first-ever reproduction of the saturation feature from QCD, and a significant success of the HAL QCD approach~\cite{Inoue:2013nfe}.
Although the obtained saturation point deviates significantly from the empirical one,
it is again primarily due to the unphysically heavy up and down quark used in our lattice QCD simulation.
Note that we have never used any phenomenological inputs for nucleon interaction, but used only QCD.
Because the lattice QCD $N\!N$ interactions are weaker than the phenomenological ones, 
the resulting binding energy is smaller than the empirical value.
From the quark mass dependence of the curve shown in the figure, 
one can expect that a result more compatible to the phenomenological (APR) one will be obtained
when we have extracted $N\!N$ interactions from QCD at the physical point.
In the figure, we see also that the saturation feature is very sensitive to change of quark mass
so that it is lost even for our second lightest quark mass case.
According to our result, it seems that the saturation feature appears again in the QCD world with very heavy quarks.

Fig.~\ref{fig:eos}, in the right panel, shows the obtained ground state energy per nucleon $E_{0}/A$
for pure neutron matter (PNM) as a function of $k_F$.
The most interesting point in the PNM EoS is the slope at large $k_F$ which determines pressure of PNM at high density.
In general, matter is more stiff when it has higher pressure.
The resulting EoS from QCD shows that PNM becomes more stiff as quark mass decreases.
It seems that QCD prediction of PNM EoS is approaching to the phenomenological (APR) one as quark mass decreases.

\begin{figure}[t]
\centering
\includegraphics[width=0.45\textwidth]{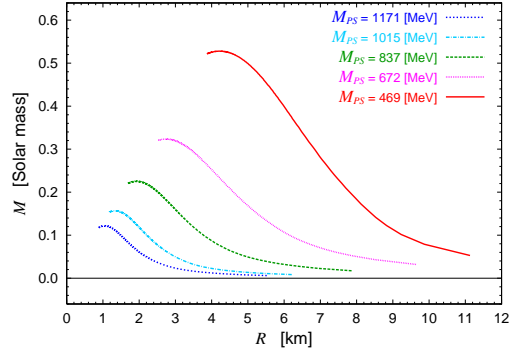}
\caption{Mass-radius relation of neutron stars obtained with
nuclear matter EoS from lattice QCD nuclear force in heavy quark region.
Crust contribution is not taken into account.}
\label{fig:neutron_star}
\end{figure}

Stiffness of PNM is very important to sustain massive neutron stars.
Fig.~\ref{fig:neutron_star} shows the mass-radius relation of neutron stars obtained with the EoS from lattice QCD nuclear force.
In this calculation, neutron-star matter consists of neutrons, protons, electrons and muons under the charge neutrality and beta equilibrium.
The Tolman-Oppenheimer-Volkoff equation~\cite{Tolman:1939jz,Oppenheimer:1939ne} is solved for spherical non-rotating neutron stars
without taking the crust into account.
We can see that the maximum neutron-star mass increases rapidly as quark mass decreases. This is due to the stiffness of PNM.
Obtained maximum mass is much smaller than the mass of already observed neutron stars.
One of the reasons of this incompatibility is the unphysically heavy up and down quark in our lattice simulations.
Hence, it is interesting to see nuclear matter EoS and mass-radius relation of neutron stars 
resulting from QCD with the physical quark mass. This is what we are going to do in the future.

%% file: section_7.tex
\section{Summary and discussion}

We have introduced our purpose and strategy in section~\ref{sec:intro}.
We want to explain or predict properties of nuclei and nuclear matter,
starting from QCD, the fundamental theory of the strong interaction.
Our strategy consists of two stages, namely we first extract the nuclear force from the lattice QCD numerical simulation,
and then we apply it to advanced few-body methods or established many-body theories.
The HAL QCD method enables us to extract potential of nuclear force from QCD on lattice.
In section~\ref{sec:hal}, we have described the method in some detail.
In particular, we discussed its advantages over the conventional method,
namely that gives a crucial solution of the plateau crisis in multi-hadron system in lattice QCD.
In section~\ref{sec:setup}, we have carried out lattice QCD numerical simulations
at five unphysical values of quark mass, 
and obtained two-nucleon potentials which possess characteristic features of phenomenological ones. 
In sections~\ref{sec:he4}, \ref{sec:medium}, and \ref{sec:matter}, we have applied the obtained potentials
to the light nucleus $^4$He, the medium-heavy nuclei $^{16}$O and $^{40}$Ca, and nuclear matter, respectively.

We have found that the nuclei $^4$He,  $^{16}$O, and $^{40}$Ca exist in
lattice QCD at a flavor $SU(3)$ point with a quark mass corresponding to the pseudo-scalar meson mass of 469 MeV.
We have deduced mass and structure of these nuclei from QCD at that quark mass. 
We have found the saturation feature of the symmetric nuclear matter at the same quark mass. 
These are certainly significant successes and progress in theoretical nuclear physics,
and demonstrates that the HAL QCD approach to nuclei from QCD is promising.

\begin{figure}[t]
\centering
\includegraphics[width=0.45\textwidth]{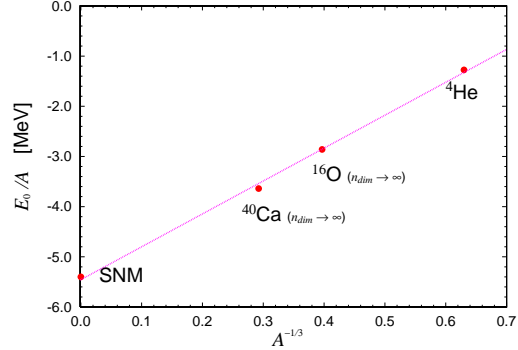}
\caption{Ground state energy per nucleon $E_0/A$ of several nucleon systems
obtained with a lattice QCD nuclear force at a quark mass corresponding to $M_{\rm PS}=469$ MeV,
as a function of $A^{-1/3}$.}
\label{fig:adepend}
\end{figure}

Fig.~\ref{fig:adepend} puts together the ground state energy per nucleon $E_0/A$
of these nucleonic systems at that quark mass, as a function of $A^{-1/3}$.
For the energy of $^{16}$O and $^{40}$Ca, a linear extrapolation to $n_{\rm dim}=\infty$
with the formula $E_0(n_{\rm dim})= E_0(\infty) + c(A) /n_{\rm dim}$, is applied.
We can see uniform $A$ dependence consistent with the Bethe-Weizs{\"a}cker mass formula
$E_0(A) = - a_{\rm V}\,A - a_{\rm S}\,A^{2/3} \cdots$,
which is known to be good for nucleon system in the real world.
Therefore, it seems that we have obtained natural nuclear system in the HAL QCD approach.

In this study, we have not considered nuclear forces in $P$, $F$, and higher partial-waves, in particular the $LS$ force.
We have checked that the odd parity forces do not have a sizable effect for the $^4$He nucleus.
However, it is known that the $LS$ force is important for the structure of nuclei, such as the magic number, 
especially at the region of heavy nuclei $A>40$.
Recently, we have developed a method to extract the odd parity nuclear force in lattice QCD simulations~\cite{Murano:2013xxa}.
It is also known that three-nucleon force is necessary for quantitative explanation of mass and structure of nuclei.
Work towards obtaining the three-nucleon force from QCD is also in progress~\cite{Doi:2011gq}.
We will include those forces in our future study on nuclei and nuclear matter from QCD.
For the medium-heavy nuclei, we have used the traditional Brueckner-Hartree-Fock theory.
In order to improve our results quantitatively,
we will use modern sophisticated theory in our future study,
such as the self-consistent Green's function method~\cite{Dickhoff:2004xx}.

We have not dealt with any hyperon interaction in this paper.
Theoretical prediction of hyperon forces based on QCD are highly desirable
since they are difficult to extract by experiment.
We can extract hyperon forces in the HAL QCD method without any fundamental difficulty~\cite{Nemura:2008sp,Inoue:2010hs}.
It is natural to consider that hyperons appear at the inner core of a neutron star.
Hence, it is interesting to study hyperon onset based on QCD in the HAL QCD approach.

We have set mass of up and down quark unphysically heavy due to the limitation of computational resources.
A lattice QCD simulation is currently under way on the K-computer at RIKEN in Japan,
to extract baryon-baryon interaction from QCD at the physical point in the HAL QCD method. 
Potentials obtained in the study will bring a new connection between QCD and nuclear physics
and astrophysics.